\documentclass[submission, Phys]{SciPost}

\hypersetup{
    colorlinks,
    linkcolor={red!50!black},
    citecolor={blue!50!black},
    urlcolor={blue!80!black}
}

\usepackage[bitstream-charter]{mathdesign}
\DeclareSymbolFont{usualmathcal}{OMS}{cmsy}{m}{n}
\DeclareSymbolFontAlphabet{\mathcal}{usualmathcal}

\usepackage[T1]{fontenc} 
\usepackage[textsize=footnotesize,textwidth=2.5cm]{todo notes}
\usepackage{cleveref}
\usepackage{caption}
\usepackage{subcaption}
\usepackage{physics}
 \def\n3{\sqrt{3}}

\def\D{\mathrm{d}}

\usepackage{colortbl}
\definecolor{LouisBlue}{RGB}{55, 114, 202}
\definecolor{LouisOrange}{RGB}{180, 54, 22}
\definecolor{LouisColor1}{RGB}{0, 118, 63}
\definecolor{LouisColor2}{RGB}{111, 73, 189}
\usetikzlibrary{decorations.markings,3d,arrows.meta}
\tikzset{middlearrow/.style={
        decoration={markings,
            mark= at position #1 with {\arrow{latex}} ,
        },
        postaction={decorate}
    }
}
\tikzset{inversemiddlearrow/.style={
        decoration={markings,
            mark= at position #1 with {-\arrow{latex[reversed]}} ,
        },
        postaction={decorate}
    }
}

\usepackage{hyperref}
\hypersetup{
    colorlinks=true,
    linkcolor=LouisBlue,
    citecolor=LouisBlue,
    urlcolor=LouisBlue,
} 

\usepackage{array}    
\usepackage{amsmath}  
\newcommand{\btikz}[1]{\begin{tikzpicture}
                          #1
                        \end{tikzpicture}}
\def\shell{\btikz{[line cap=round,scale=1.3,line width=1.5pt]
    \filldraw[gray] (0,0) circle (3pt);
    \filldraw[white](0,0)circle (1pt);
}}
\renewcommand{\bigodot}{\shell}

\begin{document}
 \hfill USTC-ICTS/PCFT-23-29\\

\begin{center}{\Large \textbf{Geometrizing the Partial Entanglement Entropy: from PEE Threads to Bit Threads}}\end{center}

\begin{center}
Jiong Lin\textsuperscript{1,2 $\ast$}, Yizhou Lu\textsuperscript{3 $\star$} and Qiang Wen\textsuperscript{4 $\dagger$}

\end{center}

\begin{center}
{\bf 1} Interdisciplinary Center for Theoretical Study, University of Science and Technology of China, Hefei, Anhui 230026, China\\
{\bf 2} Peng Huanwu Center for Fundamental Theory, Hefei, Anhui 230026, China\\
{\bf 3}  Department of Physics, Southern University of Science and Technology, Shenzhen 518055, China\\
{\bf 4 } Shing-Tung Yau Center and  School of Physics, Southeast University, Nanjing 210096, China

E-mail:
$^\ast$  {\color{blue}jionglin@ustc.edu.cn},\quad
$^\star$  {\color{blue}luyz@sustech.edu.cn},\quad 
$^\dagger$ {\color{blue}wenqiang@seu.edu.cn}

\end{center}


\section*{Abstract}
{\bf We give a scheme to geometrize the partial entanglement entropy (PEE) for holographic CFT in the context of AdS/CFT. More explicitly, given a point $\textbf{x}$ we geometrize the two-point PEEs between $\textbf{x}$ and any other points in terms of the bulk geodesics connecting these two points. We refer to these geodesics as the \textit{PEE threads}, which can be naturally regarded as the integral curves of a divergenceless vector field $V_{\textbf{x}}^{\mu}$, which we call \emph{PEE thread flow}. The norm of $V_{\textbf{x}}^{\mu}$ that characterizes the density of the PEE threads can be determined by some physical requirements of the PEE. We show that, for any static interval or spherical region $A$, a unique bit thread configuration can be generated from the PEE thread configuration determined by the state. Hence, the non-intrinsic bit threads are emergent from the intrinsic PEE threads. For static disconnected intervals, the vector fields describing a divergenceless flow is no longer suitable to reproduce the RT formula. We weight a PEE thread with the number of times it intersects with any homologous surface. Instead, the RT formula is perfectly reformulated by the minimization of the summation of PEE threads with all possible assignment of weights.
}

\vspace{10pt}
\noindent\rule{\textwidth}{1pt}
\tableofcontents\thispagestyle{fancy}
\noindent\rule{\textwidth}{1pt}
\vspace{10pt}


\section{Introduction}
In the context of AdS/CFT correspondence \cite{Maldacena:1997re}, the Ryu-Takayanagi (RT) formula \cite{Ryu:2006bv,Ryu:2006ef,Hubeny:2007xt} uncovers a relation between bulk geometry and boundary entanglement, by claiming that the entanglement entropy (EE) $S_A$ for a region $A$ in the boundary CFT is given by the area of a minimal
surface $\mathcal{E}_A$ homologous to $A$ in the AdS bulk,
\begin{equation}
    S_A=\frac{\text{Area}(\mathcal{E}_A)}{4G_N}.
\end{equation}
The RT formula was refined to its quantum corrected version, the quantum extremal surface (QES) formula \cite{Lewkowycz:2013nqa,Faulkner:2013ana,Engelhardt:2014gca}, which recently enlightens a new understanding on the black hole information paradox \cite{Hawking:1974sw,Hawking:1976ra}and was further refined towards the so called \textit{island formula} \cite{Penington:2019npb,Almheiri:2019psf,Almheiri:2019hni,Penington:2019kki,Almheiri:2019qdq}, which has recently been widely studied \cite{Marolf:2020xie,Basu:2023wmv,Lu:2022cgq,Lu:2021gmv,Basu:2022crn,KumarBasak:2020ams,Yu:2023whl,Miao:2022mdx,Li:2023fly,Geng:2020qvw,Deng:2020ent,An:2023dmo,Gan:2022jay,Hartman:2020khs,Hashimoto:2020cas,Ling:2020laa,Akal:2020twv,Wang:2021woy,Guo:2023fly,Chang:2023gkt,Afrasiar:2022fid,Geng:2023iqd}.

To give a more concrete geometric description for holographic entanglement entropy, Freedman and Headrick reinterpreted this geometric optimization problem as a flow optimization problem \cite{Freedman:2016zud,Headrick:2017ucz}\footnote{Another reformulation of the RT formula is based on the identification of minimal surfaces in Riemannian geometry through calibrations \cite{Bakhmatov:2017ihw}.}. 
Specifically, the area of the minimal surface $\mathcal{E}_A$ can be given by the 
maximum flux through the boundary region $A$ by optimizing over all possible divergenceless vector fields $v$ whose norm is upper-bounded by $|v|\leq 1/4G_N$. 
This optimized flow configuration is highly degenerate and is said to \emph{lock} the boundary region $A$.
On the minimal surface $\mathcal{E}_A$, an optimized flow should be normal to $\mathcal{E}_A$ with the norm saturating the bound $1/4G_N$. In other words, $\mathcal{E}_A$ is the bottleneck of the flow configuration.
The unoriented integration curves of this optimized flow configuration is known as \emph{bit threads} \cite{Freedman:2016zud}.
Later in \cite{Cui:2018dyq,Headrick:2020gyq}, the flow is generalized to \emph{multiflow} to simultaneously lock all the non-overlapping multi-regions.
Explicit bit thread configurations in pure AdS were first constructed by using the bulk geodesics in \cite{Agon:2018lwq}.
Then, in \cite{Agon:2020mvu}, the perturbations of bit threads configurations around pure AdS were considered by using bulk geodesics or Iyer-Wald formalism \cite{Iyer:1994ys}.
Bit threads have been also generalized to its quantum corrected version \cite{Agon:2021tia,Rolph:2021hgz} and the covariant version \cite{Headrick:2022nbe}. 
For other recent progresses on bit threads, please refer to \cite{Harper:2018sdd,Du:2019emy,Harper:2019lff,Harper:2021uuq,Lin:2022agc,Lin:2022flo,Gursoy:2023tdx}.

Despite the equivalence between bit thread description and the RT formula in computing holographic entanglement entropy, the bit thread configuration is considered to be non-intrinsic for its non-uniqueness and dependence on the choice of the region. In other words, no bit threads configuration can lock all the regions simultaneously.
For this reason, the explicit distribution of a bit thread configuration has no physical meaning.
It is very important to give particular physical meaning to the distribution of the bit threads, hence clarify the rules that uniquely determine the bit threads configurations. A natural interpretation for bit threads distribution could be the \emph{entanglement contour} $s_A(\textbf x)$ \cite{Chen:2014}, which captures the contribution from the local degree of freedom at each site $\textbf x$ to $S_A$. For a given bit thread configuration, a unique entanglement contour $s_{A}(\textbf x)$ function can be read from the configuration by computing the bit thread flux emanating from $\textbf x$ to the complement of $A$ (see \cite{Kudler-Flam:2019oru} for an explicit example). Nevertheless, even the entanglement contour is uniquely determined by the reduced density matrix $\rho_{A}$, the bit thread configuration consistent with $s_{A}(\textbf x)$ is still not unique as the thread distribution on the complement $\bar{A}$ can not be clarified. In addition, the dependence on the choice of $A$ persists.

In this paper, we will show that the \emph{partial entanglement entropy} (PEE) \cite{Wen:2018whg,Kudler-Flam:2019oru,Han:2019scu,Wen:2019iyq,Wen:2020ech,Han:2021ycp} \footnote{See \cite{Wen:2021qgx,Ageev:2021ipd,Rolph:2021nan,Camargo:2022mme,Wen:2022jxr,Lin:2022aqf,Basu:2023wmv,Lin:2023orb} for other recent progress on PEE.} can give rise to a specific and unique bit threads configuration that locks any static intervals in two dimensions and spherical regions in general dimensions in the context of AdS/CFT. Like the mutual information, the PEE measures the correlation between two regions in a certain way. The key feature of the PEE is satisfying the property of additivity. The PEE satisfies a set of physical requirements \cite{Chen:2014,Wen:2018whg}, and in Poincar\'e invariant theories the PEE can be uniquely determined by these requirements. So far, there are already several proposals to construct the PEE, including the geometric construction that works in holographic theories with a local modular Hamiltonian \cite{Wen:2018whg,Wen:2020ech,Liu:2023djf}, the additive linear combination (ALC) proposal that works in general two dimensional theories, the solution of the set of requirements, or the extensive mutual information (EMI) proposal that works for theories with conformal symmetries, and the Gaussian formula for Gaussian states of many-body systems \cite{Chen:2014,Kudler-Flam:2019nhr}. In the regimes where different proposals applies, these proposals generate highly consistent results \cite{Wen:2018whg,Wen:2020ech,Kudler-Flam:2019nhr}. The PEE encodes all the information of \emph{entanglement contour}. For a region $A$, the PEE not only characterizes the contributions from each site inside $A$, but also clarifies the the different roles played by the sites outside $A$.

Due to the property of additivity, any PEE $\mathcal{I}(A,B)$ between two non-overlapping regions $A$ and $B$ can eventually decompose into a summation of two-point PEEs,
\begin{align}
 \mathcal{I}(A,B)=\sum_{\textbf x,\textbf y}\mathcal{I}(\textbf x,\textbf y),\qquad \textbf x\in A,~~~~\textbf y\in B\,.
 \end{align} 
The two-point PEEs then fully describe the PEE structure of the state. Our scheme to geometrize the PEE is to represent the two-point PEEs as the bulk geodesics connecting these two points, which we call the \textit{PEE threads}. The PEE threads emanating from a single boundary point $\textbf x$ can be further regarded as the integral curves of a divergenceless vector field $V_{\textbf x}^{\mu}$ in the bulk, which we call the \emph{PEE thread flow}.
Then by superposing these PEE thread flows associated with all points inside the interval,we explicitly show how to get a unique bit threads configuration from the PEE threads in AdS$_3$. Interestingly, the resulting bit thread configuration respects the symmetries of the theory, and coincides with the bit threads configuration previously constructed in \cite{Agon:2018lwq}. This prescription also works for static spherical regions in higher dimensional CFTs ($d\geq$ 3). Nevertheless, this scheme fails when applied to the non-spherical boundary region in higher dimensions. We also study the PEE threads for disconnected intervals and show how PEE threads picture can interpret the phase transition of the RT surfaces in AdS$_3$/CFT$_2$. 

In Sec.\ref{sec:intro}, we will give a brief introduction to the PEE and bit threads. The scheme to geometrize the PEE is explicitly presented in Sec.\ref{sec:geo-pee}. Also, we will show how a bit thread configuration emerges from the PEE threads configuration. Explicit calculations will be carried out for intervals and spherical boundary regions in the pure AdS spacetime in general dimensions. In Sec.\ref{sec:pt-pt}, we study the reformulation of the RT formula based on the PEE thread configuration. In this case the entanglement entropy is reproduced by a summation of weighted PEE threads. The discontinuous phase transition of the RT surfaces for multi-intervals is reproduced by the switching the assignment of the weight of the PEE threads that gives the minimal value of the summation. We give a discussion in Sec.\ref{sec:con-dis}.

\section{A brief introduction to PEE and bit threads}\label{sec:intro}

\subsection{Partial entanglement entropy}
The partial entanglement entropy $\mathcal{I}(A,B)$ \cite{Wen:2018whg,Kudler-Flam:2019oru,Han:2019scu,Wen:2019iyq,Wen:2020ech,Han:2021ycp} is a measure of the correlation between two spacelike separated region $A$ and $B$. It is defined to satisfy a set of physical requirements \cite{Chen:2014,Wen:2019iyq} including  all those satisfied by the mutual information $I(A,B)$\footnote{Note that, we should not mix between the mutual information $I(A,B)$ and the PEE $\mathcal{I}(A,B)$.} and the feature of being additive. For non-overlapping regions $A$, $B$ and $C$, the physical requirements for the PEE can be briefly summarized in the following\footnote{Note that, the requirement 5 is a result of the requirements 1 and 4, hence not all of the requirements are independent. For more details about the well (or uniquely) defined scope of the PEE and the ways to construct the PEEs in different situations, the readers may consult \cite{Wen:2018whg,Kudler-Flam:2019oru,Wen:2019iyq,Han:2019scu,Wen:2020ech,Han:2021ycp,SinghaRoy:2019urc}. These details are also summarized in the background introduction sections of \cite{Camargo:2022mme,Wen:2022jxr}.}:
\begin{enumerate}
\item
\textit{Additivity:} $\mathcal{I}(A,B\cup C)=\mathcal{I}(A,B)+\mathcal{I}(A,C)$;

\item
\textit{Permutation symmetry:} $\mathcal{I}(A,B)=\mathcal{I}(B,A)$;

\item
\textit{Normalization:} $\mathcal{I}(A,\bar{A})=S_{A}$;
\item
\textit{Positivity:} $\mathcal{I}(A,B)>0$;
\item
\textit{Upper bounded:} $\mathcal{I}(A,B)\leq \text{min}\{S_{A},S_{B}\}$;
\item
\textit{$\mathcal{I}(A,B)$ should be Invariant under local unitary transformations inside $A$ or $B$};
\item
\textit{Symmetry:} For any symmetry transformation $\mathcal T$ under which $\mathcal T A = A'$ and $\mathcal T B = B'$, we have $\mathcal{I}(A,B) = \mathcal{I}(A',B')$.
\end{enumerate}

It has been shown in \cite{Casini:2008wt,Wen:2019iyq} that, the above requirements for the vacuum state of theories with Poincar\'e symmetry admit an unique solution. More interestingly, for CFTs the formula of the PEE solution can be determined up to a coefficient by imposing the above requirements except the normalization. Nevertheless, to determine the coefficient by the normalization requirement is quite subtle. Firstly, in quantum field theories the entanglement entropy of a region is divergent and depends on the explicit regularization scheme, which makes the matching between the entanglement entropy calculated by $\mathcal{I}(A,\bar{A})$ \cite{Bueno:2015rda,Bueno:2015qya,Bueno:2019mex,Bueno:2021fxb,Han:2019scu} and those calculated via other approaches quite subtle. Secondly, it is enough to determine the coefficient by imposing the normalization requirement to a connected region, for example a static spherical region. Later we will see that, when we deal with disconnected regions, for example the multi-interval cases, the solution may not exist. Our prescription is to generalize this naive normalization requirement. 

The Additivity and Permutation symmetry properties indicates that, the PEE structure is fully described by the two-point PEEs $\mathcal{I}(\textbf{x},\textbf{y})$ \cite{Wen:2019iyq}. In other words, any PEE $\mathcal{I}(A,B)$ can be evaluated by the integration (or summation for discrete systems) of certain class of two-point PEEs,
\begin{equation}
\mathcal{I}\left(A, B\right)=\int_{A} \D \sigma_{\textbf{x}} \int_{B} \D \sigma_{\textbf{y}}~ \mathcal{I}(\textbf{x}, \textbf{y}).
\end{equation}
where $\sigma_{\textbf{x}}$ and $\sigma_{\textbf{y}}$ are infinitesimal area elements located at $\textbf{x}$ and $\textbf{y}$ inside $A$ and $B$ respectively.

The entanglement contour \cite{Chen:2014} is a special type of PEE hence can also be generated from the two-point PEEs. The entanglement contour function $s_{A}(x)$ is assumed to be the density function of the entanglement entropy, which captures the contributions from the local degrees of freedom to $S_A$. Although it is hard to clarify what it means by the contribution from each degrees of freedom, it is clear that the entanglement entropy $S_{A}$ is just the collection of the contributions from all the degrees of freedom inside $A$,
\begin{equation}
    S_{A}=\int_{A}s_A(\textbf x)\D\sigma_{\textbf x}.
\end{equation}
The physical interpretation for the entanglement contour perfectly matches with the physical requirements of the PEE, which makes the PEE a natural proposal for the entanglement contour. According to the normalization property $S_{A}=\mathcal{I}(\bar{A},A)=\int_{A}\mathcal{I}(\bar{A},\textbf x )\D\sigma_{\textbf x}$, it is straightforward to propose that
\begin{align}
s_{A}(\textbf x)=\mathcal{I}(\bar{A},\textbf x)\,.
\end{align}
Similarly, the contribution from a subset $A_i$ of $A$ to $S_A$ can also be expressed as a PEE,
\begin{equation}
s_{A}(A_i)=\int_{A_i}s_A(\textbf x)\D\sigma_{\textbf x}=\mathcal{I}(\bar{A},A_i)\,.
\end{equation}

Here we only introduce one particular proposal to construct the PEE in generic two-dimensional theories with all the degrees of freedom settled in a unique order (for example settled on a line or a circle), which we call the additive linear combination (ALC) proposal \cite{Wen:2018whg,Wen:2020ech,Wen:2019iyq}. 
\begin{itemize}
\item \textbf{The ALC proposal}: 
Consider a boundary region $A$ which is partitioned in the following way, $A=\alpha_L\cup\alpha\cup\alpha_R$, where $\alpha$ is some subregion inside $A$ and $\alpha_{L}$ ($\alpha_{R}$) denotes the regions left (right) to it. The \textit{proposal} claims that:
\begin{align}\label{ECproposal}
s_{A}(\alpha)=\mathcal{I}(\alpha,\bar{A})=\frac{1}{2}\left(S_{ \alpha_L\cup\alpha}+S_{\alpha\cup \alpha_R}-S_{ \alpha_L}-S_{\alpha_R}\right)\,. 
\end{align}
\end{itemize}

Using ALC formula, the entanglement contour can be settled down. 
Then the two-point PEE can be derived by differentiating the contour function.
For a static spherical region $A=\{\textbf x| |\textbf x|^2\leq R^2\}$, the contour function for a $(d-2)$-dimensional sphere with radius $r$ is by \cite{Han:2019scu,Kudler-Flam:2019oru}\footnote{
In \cite{Han:2019scu}, this expression is derived using the ALC proposal and the trick in \cite{Casini:2011kv}.
In \cite{Kudler-Flam:2019oru}, this contour function is read from a specific construction of the bit threads \cite{Agon:2018lwq}.}
\begin{equation}\label{eq:app_contour}
s_A(r)=\frac{c}{6}\left(\frac{2 R}{R^2-r^2}\right)^{d-1}.
\end{equation}
We first determine $\mathcal{I}(0,\textbf x)$ between the origin point $r=0$ and the point $\textbf x=(r,\phi_i)$ with $r>R$.
Due to spherical symmetry, we have
\begin{equation}
    \int_{R}^{\infty} \mathcal{I}(0,r)r^{d-2}\Omega_{d-2}\D r=s_A(0)=\frac{c}{6}\frac{2^{d-1}}{R^{d-1}},
\end{equation}
where
\begin{equation}
\Omega_{d-2}=\frac{2 \pi^{\frac{d-1}{2}}}{\Gamma\left(\frac{d-1}{2}\right)}\,,
\end{equation}
is the area of the $(d-2)$-dimensional spherical surface with unit radius.
Then we have
\begin{equation}\label{I(0,x)}
    \mathcal{I}(0,\textbf x)=\frac{c}{6}\frac{2^{d-1}(d-1)}{\Omega_{d-2}|\textbf x|^{2(d-1)}}.
\end{equation}
From eq.\ \eqref{I(0,x)}, we can conclude that $\mathcal{I}(\textbf x_1,\textbf x_2)$ between two points only depends on their Euclidean distance, i.e.
\begin{equation}\label{eq:app_2p_PEE}
    \mathcal{I}(\textbf x_1,\textbf x_2)=\frac{c}{6}\frac{2^{d-1}(d-1)}{\Omega_{d-2}|\textbf x_2-\textbf x_1|^{2(d-1)}}.
\end{equation}
Taking $d=2$, we get the two-point PEE for vacuum CFT$_2$\footnote{See also the adjacency matrix defined in \cite{SinghaRoy:2019urc}.}, 
\begin{equation}\label{eq:EAM}
    \mathcal{I}(x,y)=\frac{c}{6}\frac{1}{(x-y)^2},
\end{equation}
From \eqref{eq:EAM}, we see that $\mathcal{I}(x,y)$ is proportional to the two-point function of a primary field with scaling dimension $1$, i.e. $\mathcal{I}(x,y)\propto \left\langle \mathcal{O}(x)\mathcal{O}(y)\right\rangle$, and therefore we may further relate this two-point PEE to the length $\mathcal L(x,y)$ of a bulk geodesic connecting $x$ and $y$ as $\mathcal{I}(x,y)\propto e^{-\mathcal L(x,y)/\ell_{AdS_3}}$, where $\ell_{\rm AdS_3}$ is the AdS radius.
This is an important observation that motivates us to geometrize PEE in terms of geodesics in Sec.\ref{sec:geo-pee}.

\subsection{Bit threads}
The \emph{bit threads} configuration \cite{Freedman:2016zud} is a reformulation of the RT formula to characterize the holographic entanglement entropy. 
Consider a boundary region $A$, a bit thread configuration connecting $A$ and its complement is represented by a vector flow $v_A$ that satisfies the following three properties:
\begin{enumerate}
\item $v_A$ is divergenceless;

\item the norm of $v_{A}$ is bounded by $|v_A|\leq \frac{1}{4G}$;

\item  $|v_A|= \frac{1}{4G}$ on the RT surface $\mathcal{E}_{A}$ of $A$;
\end{enumerate}
Combining the first two requirements and the max-flow min-cut (MFMC) theorem \cite{Federer:1974,Strang:1983,Nozawa:1990}, the computation of entanglement entropy of $A$ is then translated into constructing a particular $v_{A}$ with maximum flux through any codimension 2 surface $\Sigma_A$ homologous to $A$, that is
\begin{equation}\label{btEE}
    S(A)=\max\int_{A} \D \Sigma_A \sqrt{h}v^\mu_{A} n_\mu,
\end{equation}
where $h$ and $n^{\mu}$ are the induced metric and  the normal vector on $\Sigma_A$, respectively.
Intuitively, any flow from $A$ is clearly bounded by the minimal-area of bottleneck the flow has to pass through, which is just the RT surface $\mathcal{E}_{A}$, and the main content is that an optimal flow achieves this bound. As a result, a vector flow $v_A$ that optimizes the flux should satisfy the third property. Such vector flows are said to lock the region $A$ and its unoriented integral curves are known as the bit threads. 

Note that the optimal flow $v_{A}$ that satisfies the above requirements admits an enormous degeneracy and in general depends on the entangling surface we chose, hence the distribution of the bit threads has no physical meaning. Nevertheless, it will be interesting to construct explicit configurations for the flow or bit threads and endow the configuration with a physical interpretation. For example, in \cite{Agon:2018lwq} the authors considered static spherical boundary regions with radius $R$ in Poincar\'e AdS$_{d+1}$
\begin{equation}
\D s^2=\frac{1}{z^2}\left(\D r^2+r^2 \D \Omega_{d-2}^2+\D z^2\right).
\end{equation}
In their construction, the optimal flow is assumed to flow along the bulk geodesics. Regarding the divergenceless condition and the saturation of the bound $|v_A|=1/4G_N$ on $\mathcal{E}_{A}$, the vector field can be determined as 
\begin{equation}\label{bt-agon}
v^{\mu}_A=\left(\frac{2 R z}{\sqrt{\left(R^2+r^2+z^2\right)^2-4 R^2 r^2}}\right)^{d}\left(\frac{r z}{R}, \frac{R^2-r^2+z^2}{2 R}\right),
\end{equation}
where all the angular coordinate components are suppressed due to spherical symmetry. Such an optimal flow is perhaps the most natural one as it respect the symmetry of the configurations under considerations.

The relation between the entanglement contour and the bit thread configuration was first pointed out in a talk by Erik Tonni \cite{Tonni}. And in \cite{Kudler-Flam:2019oru}, the entanglement contour for static spherical regions in states dual to Poincar\'e AdS$_{d+1}$ was read from the explicit bit threads configuration \eqref{bt-agon},
\begin{equation}
s_A(r)=\frac{c}{6}\left(\frac{2 R}{R^2-r^2}\right)^{d-1}\,,
\end{equation}
where the center of the sphere is located at $r=0$. As we can see, the contour function respect the symmetry of the configuration and only depends on the radius coordinate $r$. This contour function also coincide with the one derived in \cite{Han:2019scu} based on the constant contour function measured by a Rindler observer. It is reasonable to require the bit thread configuration to reproduce the entanglement contour $s_{A}(\textbf x)$ by computing the bit thread flux that emanating from the site $\textbf x$ and anchor at the complement $\bar{A}$. Nevertheless, this additional requirement is still not enough to determine the flow configuration. 

There is no way to construct a bit thread configuration that lock all the regions we choose. It may be more realistic to take the bit thread configuration as some emergent concepts from certain intrinsic structure of the state. In the following section we will show that, the bit thread configurations that lock any static interval in holographic CFT$_2$ can be generated from the two-point PEE structure of the state, which is an intrinsic structure independent from the regions we consider.

\section{Geometrizing the PEE: from PEE threads to bit threads}\label{sec:geo-pee}

\begin{figure}
    \centering
    \includegraphics[width=0.4\textwidth]{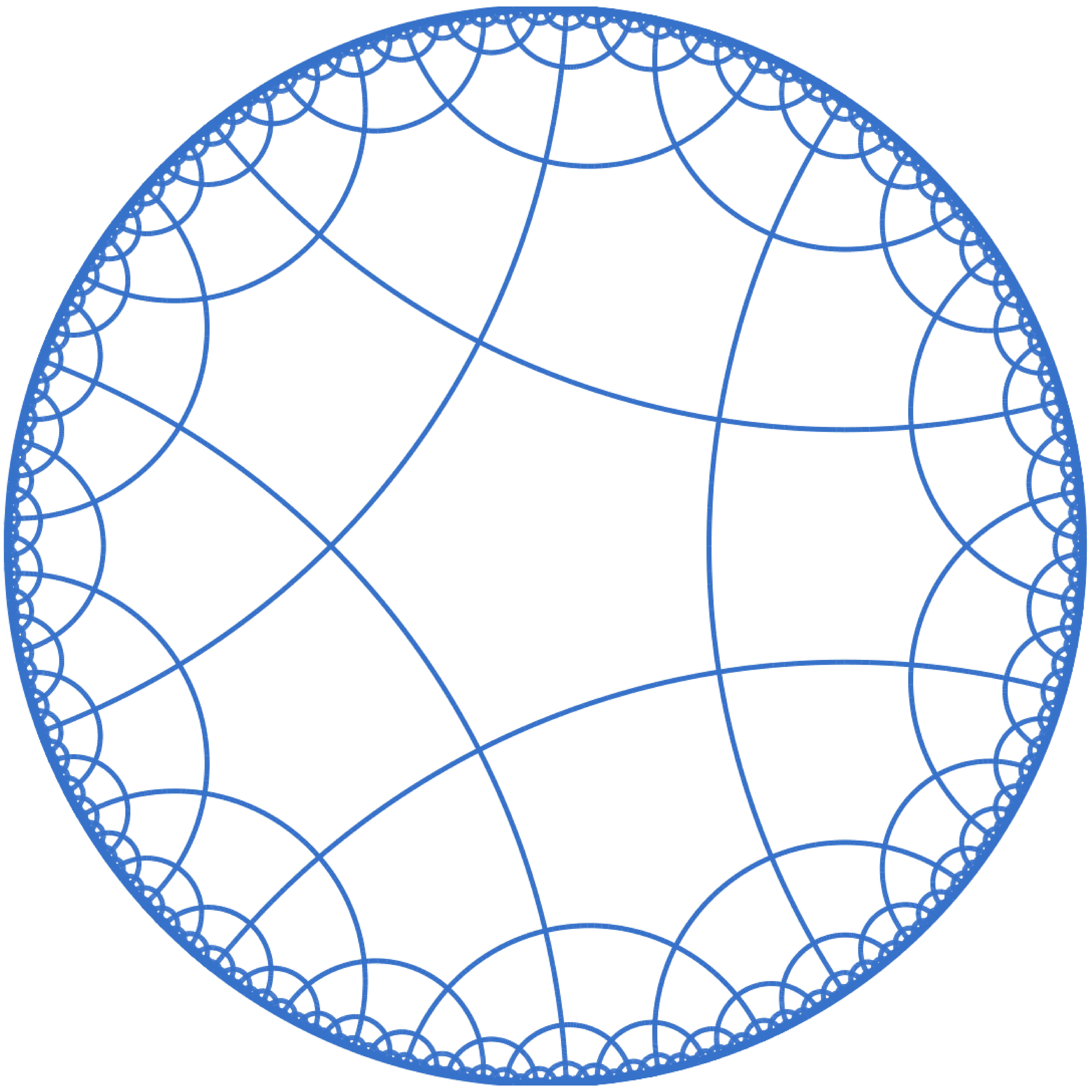}
    \caption{PEE threads on a static time slice of global AdS$_3$/CFT$_2$. }
    \label{fig:pee-thread-1}
\end{figure}

In this section, we give a scheme to geometrize the two-point PEEs with the bulk geodesics anchored on the two points. We refer these geodesics bundles as the PEE threads. The PEE threads emanating from any site $\textbf x$ can be described by a divergenceless vector fields along the geodesics emanating from $\textbf{x}$. The PEE threads emanating from different sites will intersect with each other. We will show that, the superposition of these vector fields will generate a flow in the bulk, which is just the natural optimal bit thread configuration \eqref{bt-agon} for static intervals or spherical regions in higher dimensions. 

\subsection{The scheme} \label{sec:geo-sch}
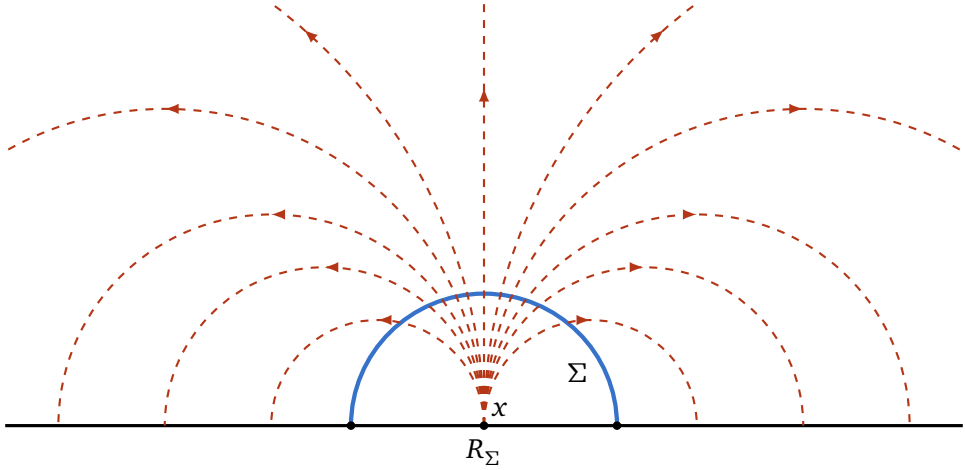
\begin{figure}
    \centering
    \begin{tikzpicture}[scale=0.7]
\clip (-9,-2) rectangle (9,8);
\draw[ultra thick,LouisBlue] (-2.5,0) arc (180:0:2.5) ;
\draw[very thick] (-9,0) -- (9,0);
\draw[] (0,-0.5) node {$R_\Sigma$};
\draw[] (1.75,1) node {$\Sigma$};
\draw[] (0.3,0.3) node {$x$};
\draw[LouisOrange,dashed,thick,middlearrow={0.8}]  (0,0) to (0,8);
\draw[dashed,thick,LouisOrange,middlearrow={0.5}] (0,0) arc (0:180:2);
\draw[dashed,thick,LouisOrange,middlearrow={0.5}] (0,0) arc (0:180:3);
\draw[dashed,thick,LouisOrange,middlearrow={0.5}] (0,0) arc (0:180:4);
\draw[dashed,thick,LouisOrange,middlearrow={0.5}] (0,0) arc (180:0:3);
\draw[dashed,thick,LouisOrange,middlearrow={0.5}] (0,0) arc (180:0:4);
\draw[dashed,thick,LouisOrange,middlearrow={0.5}] (0,0) arc (180:0:2);
\draw[dashed,thick,LouisOrange,middlearrow={0.27}] (0,0) arc (180:0:10);
\draw[dashed,thick,LouisOrange,middlearrow={0.27}] (0,0) arc (0:180:10);
\draw[dashed,thick,LouisOrange,middlearrow={0.5}] (0,0) arc (0:180:6);
\draw[dashed,thick,LouisOrange,middlearrow={0.5}] (0,0) arc (180:0:6);
\filldraw[black] (0,0) circle (2pt);
\filldraw[black] (2.5,0) circle (2pt);
\filldraw[black] (-2.5,0) circle (2pt);
\end{tikzpicture}
    \caption{The flow $V_{x}^{\mu}$ is tangent to the PEE threads (the red dashed curves).
    $\Sigma$ (the blue curve) denotes any co-dimension two surface, which is homologous to a boundary region $R_{\Sigma}$. 
    The norm of $V_{x}^{\mu}$ is determined by requiring the flux of $V_{x}^{\mu}$ on $\Sigma$ be equal to the entanglement contour $f_{R_{\Sigma}}(x)$.
 }
    \label{fig:pt-vx}
\end{figure}

Let us first consider the vacuum state of the holographic CFT$_2$ that duals to the Poincar\'e AdS$_3$. The motivation that we associate two-point PEEs with the bulk geodesics is based on an interesting observation that, the two-point PEE $\mathcal{I}(\textbf x,\textbf y)$ can be related to the length of bulk geodesic that connects two boundary points $\textbf x$ and $\textbf y$.
Specifically, by inserting the geodesic length
\begin{equation}
    \mathcal L(\textbf x,\textbf y)=\ell_{\rm AdS_3}\log\frac{(\textbf x-\textbf y)^2}{\delta^2},
\end{equation}
into \eqref{eq:EAM}, we have 
\begin{equation}
    \mathcal{I}(\textbf x,\textbf y)=\frac{\ell_{\rm AdS_3}}{4G_N \delta^2}e^{-\mathcal L(\textbf x,\textbf y)/\ell_{\rm AdS_3}},
\end{equation}
where $\delta$ is the UV cutoff.
Also, the geodesic is the most natural geometric object connecting two boundary points. Hence, it is a good starting point to geometrize the two-point PEEs in terms of their corresponding bulk geodesics. We name these geodesics bundles as the \emph{PEE threads}. In Fig.\ref{fig:pee-thread-1}, we illustrate the PEE threads on a static time slice of global AdS$_3$.  

For any boundary point $\textbf x$, the PEE threads emanating from it can be understood as the integral curves of a divergenceless vector field $V^{\mu}_{\textbf x}$, and the norm of the vector field characterizes the density of the PEE threads (see Fig.\ref{fig:pt-vx} for an illustration). 
We refer to $V^{\mu}_{\textbf x}$ as the \emph{PEE thread flow} vector field, which can be written as
\begin{align}
 V_{\textbf x}^\mu=|V_{\textbf x}|\tau_{\textbf x}^\mu\,,
 \end{align} 
where $\tau_{\textbf x}^\mu$ represents the tangent unit vectors of the PEE threads emanating from $\textbf x$. Our main task is to determine the norm $|V_{\textbf x}|$ of this vector field. Since the entanglement entropy is a collection of the two-point PEEs, it is reasonable to require that the contribution for site $\textbf x$ to $S_{A}$, or the value of $s_{A}(\textbf x)$, should be captured by the number of the PEE threads that connecting $\textbf x$ and $\bar{A}$, i.e., the flux of the PEE thread flow $V_{\textbf x}^\mu$ on any co-dimension two bulk hypersurface $\Sigma$ homologous to a boundary region $A$.
    \begin{equation}\label{rela-norm}
       \text{Requirement:}\quad \int_{\bar A} \D \sigma_{\textbf y}~ \mathcal{I}(\textbf x,\textbf y)=s_{A}(\textbf x)=\int_{\Sigma}\D \Sigma~\sqrt{h}V_{\textbf x}^\mu n_\mu,
    \end{equation}
where $n^{\mu}$ is the unit vector normal to $\Sigma$. Since the PEE thread flow is divergenceless, the flux is independent from the choice of the co-dimension two homologous surface. 

Note that, our scheme to geometrize the two-point PEEs as bulk geodesics gives us a geometric picture for the fine structure of entanglement, hence contains much more information than the above requirement. For any point $\textbf x$ inside $A$, the PEE threads emanating from $\textbf x$ give a one-to-one mapping between the points on $\Sigma$ and the points on $\bar{A}$. Indeed, the differential version of the requirement \eqref{rela-norm} can fully characterize the feature of our scheme, which is just
\begin{align}\label{drela-norm}
\D \sigma_{\textbf y} ~\mathcal{I}(\textbf x,\textbf y)=\D \Sigma~\sqrt{h}V_{\textbf x}^\mu n_\mu,
\end{align}
where the area element $\D{\Sigma}$ on $\Sigma$ is mapped to the area element $\D\sigma_{\textbf y}$ on $\bar{A}$ via the one-to-one mapping determined by the PEE threads (See Fig.\ref{diff}). We will see that, when choosing an proper $\Sigma$, the norm $|V_{\textbf x}|$ of the vector field can be easily settled down by the above requirement \eqref{drela-norm}.

\begin{figure}
\centering
\begin{tikzpicture}[scale=1]
    \coordinate [label=below:$x$] (A) at (0,0);
    \draw[very thick] (-3,0) -- (5,0);
    \draw[ultra thick,LouisBlue] (-2.5,0) arc (180:0:2.5) ;
    \draw[LouisOrange,middlearrow={0.66},dashed,thick](0,0) arc(180:0:2.1);
    \filldraw [LouisOrange] (53.4704:2.5) circle (1.5pt);
    \draw[LouisOrange,middlearrow={0.77},dashed,thick](0,0) arc(180:0:1.5);
    \filldraw [LouisOrange] (33.5573:2.5) circle (1.5pt);
    \draw [] (3.6,-0.8) node[label={[scale=1]$\mathrm d \sigma_y$}]{};
    \draw [] (33:2.6) node[label={[scale=1]$\mathrm d \Sigma$}]{};
    \filldraw[black] (0,0) circle (1.5pt);
    \filldraw[black] (3,0) circle (1.5pt);
    \filldraw[black] (4.2,0) circle (1.5pt);
\end{tikzpicture}
\caption{For a fixed point $x$ in $A$, the PEE threads give a one-to-one mapping from $\D \Sigma$ on $\Sigma$ to $\D\sigma_y$ on boundary.  }
\label{diff}
\end{figure}
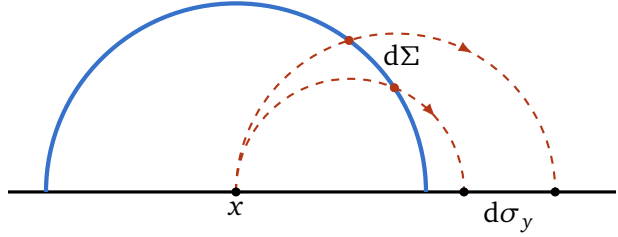

\begin{figure}
\hspace*{-20pt}
\begin{tikzpicture}[scale=0.6]
\clip (-7,-1) rectangle (7,6);
\draw[ultra thick,LouisBlue] (-4,0) arc (180:0:4) ;
\draw[very thick] (-9,0) -- (9,0);
\draw[] (0,-0.4) node {$A$};
\draw[] (-4,2) node {$\mathcal E_A$};
\draw[LouisOrange,thick,dashed,middlearrow={0.2},middlearrow={0.6}] (2,0) arc (0:180:5);
\draw[LouisOrange,thick,dashed,middlearrow={0.2},middlearrow={0.6}] (-2,0) arc (180:0:5);


\draw[LouisOrange,thick,dashed,middlearrow={0.2},middlearrow={0.6}] (3.2,0) arc (0:180:4.1);
\draw[LouisOrange,thick,dashed,middlearrow={0.2},middlearrow={0.6}] (-3.2,0) arc (180:0:4.1);
\draw[LouisOrange,thick,dashed,middlearrow={0.2},middlearrow={0.6}] (0,0)--(0,9);
\filldraw[black] (0,4) circle (2pt);
\filldraw[black] (0,0) circle (2pt);
\filldraw[black] (-4,0) circle (2pt);
\filldraw[black] (4,0) circle (2pt);
\draw(0.5,4.5) node [fill=white,scale=0.7]{};
\draw(0.5,4.5) node [scale=0.8]{$Q$};
\end{tikzpicture}
\begin{tikzpicture}[scale=0.6]
\clip (-7,-1) rectangle (7,6);
\draw[ultra thick,LouisBlue] (-4,0) arc (180:0:4) ;
\draw[very thick] (-9,0) -- (9,0);
\draw[] (0,-0.4) node {$A$};
\draw[] (-4,2) node {$\mathcal E_A$};
\draw[LouisColor1,thick,dashed,middlearrow={0.6}] (-3,0) arc (180:90:3);
\draw[LouisColor1,thick,dashed,middlearrow={0.6}] (3,0) arc (0:90:3);
\draw[LouisOrange,thick,dashed,middlearrow={0.2},middlearrow={0.6}] (1.35425,0) arc (0:180:4);
\draw[LouisOrange,thick,dashed,middlearrow={0.2},middlearrow={0.6}] (-1.35425,0) arc (180:0:4);
\draw[LouisOrange,thick,dashed,middlearrow={0.2},middlearrow={0.6}] (0,0)--(0,9);
\filldraw[black] (0,3) circle (2pt);
\filldraw[black] (0,0) circle (2pt);
\filldraw[black] (-4,0) circle (2pt);
\filldraw[black] (4,0) circle (2pt);
\draw(0.5,3.3) node [fill=white,scale=0.7]{};
\draw(0.5,3.3) node [scale=0.8]{$Q$};
\end{tikzpicture}
    \caption{Superposition of the PEE thread flow in Poincar\'e AdS with a spherical boundary region $A$.
    For the field point $Q$ on the RT surface $\mathcal E_A$ (see the left figure), only \emph{outer threads} pass through $Q$. 
    For $Q$ inside the entanglement wedge of $A$ (see the right figure), the \emph{inner threads} (green dashed curves) also pass through $Q$.
    Physically, the \emph{bit thread flow} $V_A^{\mu}$ should only count the contributions from \emph{outer threads}.
    However, since the \emph{inner threads} have the zero net contributions to $V_A^{\mu}$, it is safe to integrate over both outer and \emph{inner threads} to obtain $V_A^{\mu}$.
    }
    \label{fig:pt-sp}
\end{figure}
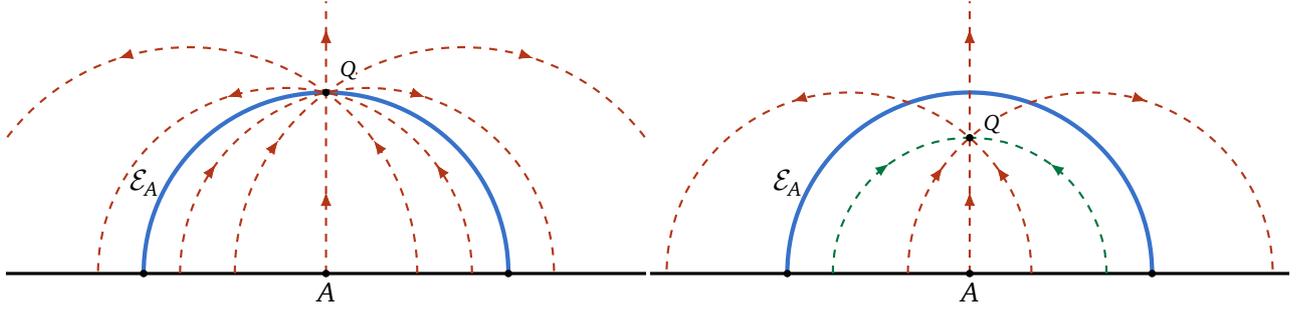

The PEE threads emanating from different boundary points could intersect with each other. Since the evaluation of the entanglement entropy $S_{A}$ is equivalent to counting all the PEE threads coming out from $A$, it is natural to consider the superposition of all the PEE thread flows $V_{\textbf x}$ with $\textbf x$ inside $A$. This results in a divergenceless vector field $V_{A}$ which we call the \emph{bit thread flow} vector field. 
Soon, we will see that such a \emph{bit thread flow} satisfies all the requirements for the vector field representing the bit threads, and thus the name.

Given a connected boundary region $A$, the PEE threads can always be classified into two classes:
\begin{itemize}
    \item 
    the \emph {inner threads} of $A$ that emanate and terminate inside $A$,
    \item 
    the \emph {outer threads} of $A$ that emanate inside $A$ and terminate outside $A$.
\end{itemize}
Only the \emph{outer threads} of $A$ (see Fig.\ref{fig:pt-sp}) contribute to the entanglement entropy. One may intend to construct a vector field, which is the superposition of all the \emph{outer threads}. Nevertheless, this is not necessary since that, for any \emph{inner thread} representing $\mathcal{I}(\textbf x, \textbf y)$, the PEE thread will be counted twice (multiplied by corresponding integral measures) with inverse directions (see the green dashed curves in the right of Fig.\ \ref{fig:pt-sp} for an illustration). According to the permutation property $\mathcal{I}(\textbf x, \textbf y)=\mathcal{I}(\textbf y,\textbf x)$ of the PEE and the divergenceless property of the \emph{PEE thread flow}, the \emph{inner threads} cancel with each other under superposition. Therefore, there is no need to distinguish between the \emph{inner} and \emph{outer threads} and the \emph{bit thread flow} $V_A^{\mu}$ can be achieved by integrating all the \emph{PEE thread flow} $V_{\textbf x}$ with $\textbf x \in A$, i.e.
\begin{equation}
       V_A^\mu= \int_{\text{outer threads of $A$} }\D^{d-1} \textbf x V_{\textbf x}^\mu
       =\int_{\text{all threads of $A$} }\D^{d-1} \textbf x V_{\textbf x}^\mu.
    \end{equation} 

According to the requirement \eqref{rela-norm}, since the flux of the \emph{PEE thread flow} $V_{\textbf x}$ gives the entanglement contour $s_{A}(\textbf x)$, the flux of the \emph{bit thread flow} $V_A^{\mu}$ should recover the entanglement entropy $S_{A}$. Soon we will check that, for static intervals or spherical regions the inequality $|V_A|\leq 1/4G_N$ is satisfied by $V_A^{\mu}$ inside the entanglement wedge, and is saturated only on the RT surface. Together with the divergenceless property of $V_{A}$, the requirements for the vector field to describe bit threads are all satisfied by $V_{A}$.

In the next sub-sections, we will explicitly construct vector fields $V_{A}$ for static intervals and spherical regions from the PEE structure and show that $|V_{A}|=\frac{1}{4G}$ is satisfied on the RT surface $\mathcal{E}_{A}$.

\subsection{PEE threads and bit threads in AdS$_3$}
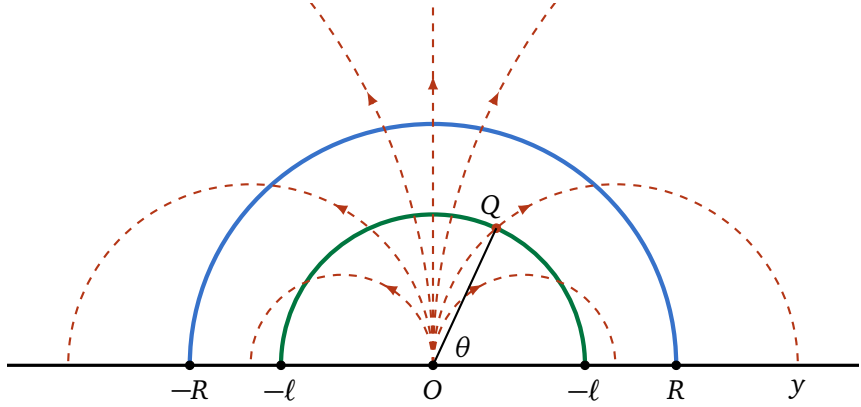
\begin{figure}
    \centering
    \begin{tikzpicture}[scale=0.8]
\clip (-7,-1) rectangle (7,6);
\draw[ultra thick,LouisBlue] (-4,0) arc (180:0:4) ;
\draw[ultra thick,LouisColor1] (-2.5,0) arc (180:0:2.5) ;
\draw[very thick] (-9,0) -- (9,0);
\draw[] (0,-0.4) node {$O$};
\draw[] (-2.5,-0.4) node {$-\ell$};
\draw[] (2.5,-0.4) node {$-\ell$};
\draw[] (-4,-0.4) node {$-R$};
\draw[] (4,-0.4) node {$R$};
\draw[] (6,-0.4) node {$y$};
\draw[LouisOrange,dashed,thick,middlearrow={0.6}]  (0,0) to (0,8);
\draw[dashed,thick,LouisOrange,middlearrow={0.35}] (0,0) arc (0:180:3);
\draw[dashed,thick,LouisOrange,middlearrow={0.35}] (0,0) arc (180:0:3);
\draw[dashed,thick,LouisOrange,middlearrow={0.35}] (0,0) arc (0:180:1.5);
\draw[dashed,thick,LouisOrange,middlearrow={0.35}] (0,0) arc (180:0:1.5);
\draw[dashed,thick,LouisOrange,middlearrow={0.15}] (0,0) arc (180:0:10);
\draw[dashed,thick,LouisOrange,middlearrow={0.15}] (0,0) arc (0:180:10);
\filldraw[black] (0,0) circle (2pt);
\filldraw[black] (2.5,0) circle (2pt);
\filldraw[black] (-2.5,0) circle (2pt);
\filldraw[black] (-4,0) circle (2pt);
\filldraw[black] (4,0) circle (2pt);
\filldraw[LouisOrange] (65.3757:2.5) circle (2pt);
\draw[black,thick] (0,0) -- (65.3757:2.5);
\draw[] (70:2.8) node {$Q$};
\draw[] (0.5,0.3) node {$\theta$};
\end{tikzpicture}
    \caption{Determine the \emph{PEE thread flow} $V^{\mu}_{O}$ at the point $Q=(\bar r,\bar z)$ by letting the flux of the PEE threads (the red curves) through the reference RT surface $\Sigma$ (the green semicircle) with the radius $\ell=\sqrt{\bar r^2+\bar z^2}$ being equal to the entanglement contour of $r=0$ to the region $[-\ell,\ell]$.
    }
    \label{fig:pt-v}
\end{figure}

Now we explicitly construct the \emph{PEE thread flow} $V_{r}^\mu$ and the \emph{bit thread flow} $V_A^\mu$ in AdS$_3$. The Poincar\'e metric on a static time slice is given by
\begin{equation}
    \D s^2=\frac{1}{z^2}(\D r^2+\D z^2).
\end{equation}
We first determine $V_{O}^{\mu}$ which describes the PEE threads emanating from the origin point $O$ settled at $r=0$, then generalize $V_{O}^{\mu}$ to arbitrary boundary point $V_{r}^{\mu}$ under a translation in the $r$ direction.

For an arbitrary bulk point $Q=(\bar r,\bar z)=\ell(\cos\theta,\sin\theta)$ with  
\begin{equation}
    \ell=\sqrt{\bar r^2+\bar z^2},\quad 
    \tan\theta=\frac{\bar z}{\bar r},
\end{equation}
let us first determine the direction of $V_{O}^{\mu}$, which is the tangent unit vector of the PEE thread (or geodesic) connecting $O$ and $Q$. Such a thread will anchor at the boundary on another point $(r,z)=(y,0)$ (see Fig.\ref{fig:pt-v}), where $y$ can be determined, 
\begin{equation}\label{solu-y}
    (\ell\cos\theta-y/2)^2+\ell^2\sin^2\theta=y^2/4,\quad \Rightarrow \quad y=\frac{\ell}{\cos\theta}.
\end{equation}
Then the tangent unit vector of the PEE thread is given by
\begin{equation}
    \tau_{O}^\mu(Q)=\frac{\ell\sin\theta}{y/2}(\ell\sin\theta,-\ell\cos\theta+y/2)=\frac{2\bar z\bar r}{\bar r^2+\bar z^2}\left(
    \bar z,\frac{\bar z^2-\bar r^2}{2\bar r}
    \right),
\end{equation}
and the \emph{PEE thread flow} can be written as 
\begin{equation}\label{vx=0-direc}
    V_{O}^\mu(Q)=|V_{O}(Q)|\tau_{O}^\mu(Q)=
    \frac{2\bar z\bar r|V_{O}(Q)|}{\bar r^2+\bar z^2}\left(
    \bar z,\frac{\bar z^2-\bar r^2}{2\bar r}
    \right).
\end{equation}

To further determine its norm, we should make use of the requirement for the flow \eqref{rela-norm}.
Let us choose a reference surface $\Sigma$ (which is the RT surface for $[-\ell,\ell]$ and is green in Fig.\ref{fig:pt-v}) that passes the point $Q$ as 
\begin{equation}
   \Sigma:  r^2+ z^2=\ell^2,
\end{equation}
with its normal unit vector at the point $Q$ given by
\begin{equation} n_{\Sigma}^\mu(Q)=\ell\sin\theta(\cos\theta,\sin\theta).
\end{equation}
The flux of the vector field $V_{O}^{\mu}$ through the RT surface $\Sigma$ is given by
\begin{equation}
\begin{aligned}
    \text{Flux}(V_{O}^{\mu},\Sigma)=&\int_{\Sigma} \D\theta \sqrt{h_{\theta\theta}} V_{O}^\mu(\theta) n_{\Sigma,\mu}\\
    =&
     \int_{\Sigma}  \D \theta \sqrt{h_{\theta\theta}}|V_{O}(\theta)|\sin\theta,
\end{aligned}
\end{equation}
where $h_{\theta\theta}=1/\sin^2\theta$ is the $\theta\theta$-component of the induced metric on $\Sigma$, and $V_{O}^\mu(\theta)$ is the value of the vector field $V_O$ on $\Sigma$ parameterized by $\theta$.

According to the requirement \eqref{rela-norm}, the flux equation equals to collecting all the PEE threads that emanate from $O$ and terminate on the complement $(-\infty,-\ell)\cup(\ell,\infty)$ region, i.e.
\begin{equation}\label{dvc}
    \int_{0}^{\pi} \D \theta~ \sqrt{h_{\theta\theta}}|V_{O}(\theta)|\sin\theta
    =\int_{(-\infty,-\ell)\cup(\ell,\infty)} \D y~\mathcal{I}(0,y).
\end{equation}
We can also parameterize $\Sigma$ using coordinate $r=y$ of the point where the outer PEE threads terminate, which is related to the parameter $\theta$ via the relation \eqref{solu-y}, i.e. $y=R/\cos\theta$. This is just the one-to-one mapping between the points on $\Sigma$ and points on $\bar{A}$ we mentioned near \eqref{drela-norm}. Then we arrive at the following equation for any $\ell$,
\begin{equation}\label{eq:dy_equality}
    \int_{(-\infty,-\ell)\cup(\ell,\infty)} \D y~\frac{\ell |V_{O}(\theta)|}{y^2\sin\theta}=\int_{(-\infty,-\ell)\cup(\ell,\infty)} \D y~ \mathcal{I}(0,y).
\end{equation}
where the two-point PEE is given by $\mathcal{I}(0,y)={1}/{(4G_N y^2)}$ following \eqref{eq:EAM}.
Then the requirement \eqref{drela-norm} indicates that,
\begin{equation}\label{eq:|V_x|}
    |V_{O}(Q)|=\frac{\mathcal{I}(0,y)y^2\sin\theta}{\ell}=\frac{1}{4G_N}\frac{\bar z}{\bar r^2+\bar z^2},
\end{equation}
where we have written $\theta$ and $\ell$ in terms of $\bar{z}$ and $\bar{r}$. With the norm of the \emph{PEE thread flow} determined, it is now straightforward to write down the explicit formula for the \emph{PEE thread flow},
\begin{equation}\label{vx=00}
    V_{O}^\mu(Q)=\frac{1}{4G_N}\frac{2\bar z^2\bar r}{(\bar r^2+\bar z^2)^2}\left(\bar z,\frac{\bar z^2-\bar r^2}{2\bar r}\right).
\end{equation}
 
Due to shift symmetry along $r$-direction,
the \emph{PEE thread flow} with $r=r_0$ can be obtained by replacing $\bar r$ with $\bar r-r_0$ in \eqref{vx=00} and  
\begin{equation}\label{vx=0}
    V_{r_0}^\mu(Q)=\frac{1}{4G_N}\frac{2\bar z^2(\bar r-r_0)}{((\bar r-r_0)^2+\bar z^2)^2}\left(z,\frac{\bar z^2-(\bar r-r_0)^2}{2(\bar r-r_0)}\right).
\end{equation}
The above expression for the PEE threads from $x$ is of the central importance of this paper.
One can check that the vector field \eqref{vx=0} is divergenceless.
 
Then we turn to the \emph{bit thread flow} $V_A$, which is the summation (or superposition) of all the \emph{PEE thread flow} $V_{r_0}^{\mu}$ emanating from the points inside an interval $A=[-R,R]$. The computation is just a simple integration of $V_{r_0}$ over the interval $[-R,R]$,
\begin{align}
    V_A^\mu(Q)&
    =\int^{R}_{-R}\D r_0V_{r_0}^{\mu}=\frac{\bar z^2}{4G_N}\frac{2R}{((R-\bar r)^2+\bar z^2)((R+\bar r)^2+\bar z^2)}
    \left(2\bar z \bar r,R^2-\bar r^2+\bar z^2\right).\label{eq:2d,vaz}
\end{align}

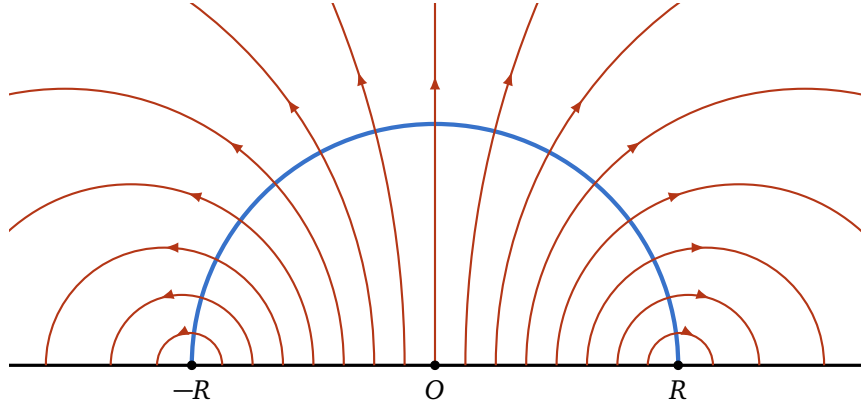
\begin{figure}
    \centering
    \begin{tikzpicture}[scale=0.8]
\clip (-7,-1) rectangle (7,6);
\draw[ultra thick,LouisBlue] (-4,0) arc (180:0:4) ;
\draw[very thick] (-9,0) -- (9,0);
\draw[] (0,-0.4) node {$O$};
\draw[] (-4,-0.4) node {$-R$};
\draw[] (4,-0.4) node {$R$};
\draw[LouisOrange,thick,middlearrow={0.6}]  (0,0) to (0,8);
\draw[LouisOrange,thick,middlearrow={0.2}]  (1,0) arc (180:0:7.5);
\draw[LouisOrange,thick,middlearrow={0.1}]  (0.5,0) arc (180:0:15.75);
\draw[LouisOrange,thick,middlearrow={0.3}]  (1.5,0) arc (180:0:4.58333);
\draw[LouisOrange,thick,middlearrow={0.4}]  (2,0) arc (180:0:3);
\draw[LouisOrange,thick,middlearrow={0.5}]  (2.5,0) arc (180:0:1.95);
\draw[LouisOrange,thick,middlearrow={0.65}]  (3.5,0) arc (180:0:0.535714);
\draw[LouisOrange,thick,middlearrow={0.6}]  (3.,0) arc (180:0:7/6);
\draw[LouisOrange,thick,middlearrow={0.2}]  (-1,0) arc (0:180:7.5);
\draw[LouisOrange,thick,middlearrow={0.1}]  (-0.5,0) arc (0:180:15.75);
\draw[LouisOrange,thick,middlearrow={0.3}]  (-1.5,0) arc (0:180:4.58333);
\draw[LouisOrange,thick,middlearrow={0.4}]  (-2,0) arc (0:180:3);
\draw[LouisOrange,thick,middlearrow={0.5}]  (-2.5,0) arc (0:180:1.95);
\draw[LouisOrange,thick,middlearrow={0.65}]  (-3.5,0) arc (0:180:0.535714);
\draw[LouisOrange,thick,middlearrow={0.6}]  (-3.,0) arc (0:180:7/6);
\filldraw[black] (0,0) circle (2pt);
\filldraw[black] (-4,0) circle (2pt);
\filldraw[black] (4,0) circle (2pt);
\end{tikzpicture}
    \caption{Bit threads configuration (the red curve) in pure AdS$_3$ constructed from PEE threads. 
    This bit threads configuration is the integral curves of the \emph{bit thread flow} $V^\mu_A$.
    They are the geodesics normal to the RT surface of $A$.
    }
    \label{fig:pt-bt}
\end{figure}

With the explicit formula for the \emph{bit thread flow} $V_{A}$ given, it is straightforward to check that $V_{A}$ satisfies all the three requirements for it to describe bit threads. Firstly, since the \emph{PEE thread flow} $V_{r_0}$ is divergenceless, $V_A$ should also be divergenceless. Secondly, if we set $(\bar{r},\bar{z})=(R\cos \theta,R\sin \theta)$ hence $Q$ is on the RT surface, we find that
\begin{align}
 V_A^{\mu}(Q)=\frac{R\sin\theta}{4G_N}(\cos\theta,\sin\theta) \,,
 \end{align}
where we can read $|V_A(Q)|=1/(4G_N)$ and $V_A(Q)$ is normal to the RT surface. At last, we can calculate $|V_A|$ anywhere in the bulk which is given by
\begin{align}
|V_A|=\frac{2 R \bar{z}}{\sqrt{\left(R^2+\bar{r}^2+\bar{z}^2\right)^2-4 R^2 \bar{r}^2}}.
\end{align}
For any bulk point $(\bar{r},\bar{z})=l(\cos \theta,\sin\theta)$ away from the RT surface, i.e. $l\neq R$, we have $\left(R^2+\bar r^2+\bar z^2\right)^2-4 R^2 \bar r^2>4R^2l^2 \sin^2\theta$, which indicates that $|V_A|<1/(4G_N)$ at the points away from the RT surface. 

In fact, according to the PEE threads picture, that $|V_A|<1/4G_N$ away from RT surface is manifest, and we have the following statement:
\begin{itemize}
 \item 
Provided that, for any region $A$ the \emph{bit thread flow} $V_A^{\mu}$ is normal to the RT surface and satisfies $|V_A|=1/4G$ on the RT surface, then we have $|V_A|<1/4G$ away from RT surface.
\end{itemize}
We leave the proof of this statement in Appendix \ref{app:pos}.
With these requirements satisfied, it is sufficient to say that, the \emph{bit thread flow} gives an optimal bit thread configuration.

In Fig.\ref{fig:pt-bt}, we plot this bit threads configuration using the integral curves of the \emph{bit thread flow} $V^\mu_A$, and find that they are just the bulk geodesics normal to the RT surface $\mathcal{E}_A$. This is not surprising as we can check that, the \emph{bit thread flow} $V_A^{\mu}$ \eqref{eq:2d,vaz} we constructed coincides exactly with the one found in \cite{Agon:2018lwq} in $d=2$. As we have mentioned, this is perhaps the most natural bit thread configuration that respect the symmetries of the configuration. In the following subsection, we will show that the coincidence also happens in higher dimensions.

\subsection{PEE threads and bit threads in AdS$_{d+1}$}
For the vacuum CFT$_d$ on a static time slice, the two-point PEE is given by
\begin{equation}\label{Ixy-generalD}
    \mathcal{I}(\textbf x_1,\textbf x_2)=\frac{c}{6}\frac{2^{d-1}(d-1)}{\Omega_{d-2}|\textbf x_2-\textbf x_1|^{2(d-1)}}.
\end{equation}
The metric of the dual Poincar\'e AdS$_{d+1}$ at a time slice is given by
\begin{equation}
\D s^2=\frac{1}{z^2}\left(\D \textbf{x}^2+\D z^2\right)=\frac{1}{z^2}\left(\D r^2+r^2 \D \Omega_{d-2}^2+\D z^2\right),
\end{equation}
where $\textbf x=\left(x_1,x_2,\cdots, x_{d-1}\right)$ and 
\begin{align}
\D \Omega_{d-2}^2=\D\phi_1^2+\sin^2\phi_1\D\phi_2^2+\cdots +\sin^2\phi_1\cdots \sin^2\phi_{d-3}\D\phi_{d-2}^2.
\end{align}

Again, we first consider the \emph{PEE thread flow} $V_{O}$ for the origin $O$ at $r=0$. Due to the rotational symmetry of $V_{O}$, we will restrict to the 2-dimensional slice with $\phi_i=0$ and employ the similar strategy as in case of AdS$_{3}$. This simplifies the analysis since the bulk geodesics have the same function as those in AdS$_3$ at this slice. For a bulk point $Q=(\bar r,\bar z)=(\ell\cos\theta,\ell\sin\theta)$ on the $\phi_i=0$ slice, the \emph{PEE thread flow} is tangent to the PEE threads, hence
\begin{equation}
    V_{O}^\mu(Q)=
    \frac{2\bar z\bar r|V_{O}(Q)|}{\bar r^2+\bar z^2}\left(
    \bar z,\frac{\bar z^2-\bar r^2}{2\bar r}
    \right).
\end{equation}
The flux equation \eqref{rela-norm} becomes 
\begin{equation}
    \int_\Sigma \D\theta~ \frac{\bar r^{d-2}}{\bar z^{d-2}} |V_{O}|=\int \D y~ y^{d-2}\mathcal{I}(0,y),
\end{equation}
where $\Sigma$ is a reference surface $ r^2+z^2=\ell^2$ that passes through $Q$.
Since the PEE threads have the same function as those in AdS$_3$, the one-to-one mapping between the points on $\Sigma$ and the boundary points outside the region associated to $\Sigma$ on the $\phi_i=0$ slice is again given by the relation \eqref{solu-y}, i.e. $y=\ell/\cos\theta$. Then the norm of the \emph{PEE thread flow} is given by
\begin{equation}
    |V_{O}(Q)|=\frac{1}{4G_N}\frac{2^{d-1}(d-1)}{\Omega_{d-2}}\frac{\bar z^{d-1}}{(\bar r^2+\bar z^2)^{d-1}}.
\end{equation}
And then the \emph{PEE thread flow} is given by
\begin{equation}\label{vx=0-HD}
    V_{O}^\mu(Q)=\frac{2^d \bar z^d}{4G_N}\frac{(d-1)}{\Omega_{d-2}}\frac{\bar r}{(\bar r^2+\bar z^2)^d}
    \left(
    \bar z,\frac{\bar z^2-\bar r^2}{2\bar r}
    \right).
\end{equation}
Note that, on the $\phi_i=0$ slice the $V_{O}^{\phi_i}$ components are zero.

When $d\geq 3$, the explicit formula of the vector field $V_{P}$ for an arbitrary boundary point $P$ is much more complicated. Nevertheless, due to the symmetry of the configuration, we will see that solving $V_{O}^\mu$ is enough to determine the \emph{bit thread flow} $V_A^{\mu}$ for any static spherical region $A$. For any point $P:(\textbf x,0)$ on the boundary and $Q:(\textbf y,\bar{z})$ in the bulk, one can consider the projection point $Q_b:(\textbf y,0)$ of $Q$ on the boundary. Due to the translational and rotational symmetries of $V_{P}$, it will be useful to note that, $V_{P}(Q)$ takes the same formula as \eqref{vx=0-HD}, with $\bar{r}$ replaced by $|\textbf x -\textbf y|$ and the $r$ coordinate parameterizing the direction from $P$ to $Q$.  

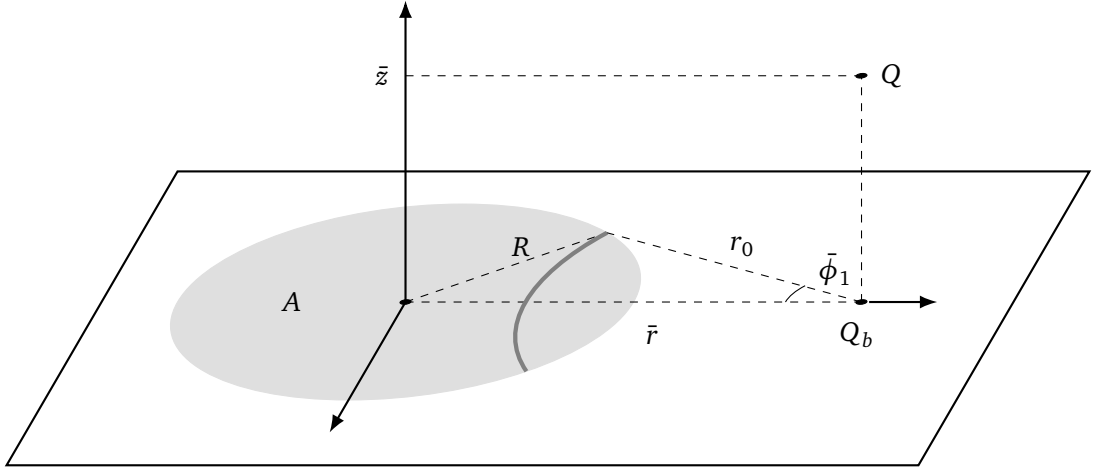
\begin{figure}
    \centering
    \begin{tikzpicture}[z={(0:10mm)},x={(-120:5mm)}]]
    \begin{scope}[canvas is xz plane at y=0]
        \filldraw[lightgray!50] (0,0) circle [radius=3];
    \filldraw[black] (0,0) circle (2pt);
    \filldraw [black] (0,6) circle (2pt);
    \draw[dashed] (0,0) -- (0,6);
    \draw[dashed] (0,0) -- (135:3);
    \draw[gray,ultra thick](-2.12132,2.12132) arc (151.325+90:208.675+90:4.42088); 
    \draw[black] (0,5) arc (-90:-120:1);
    \draw[dashed](0,6)--(-2.12132,2.12132);
    \draw[] (-1,5.4) node{$\bar\phi_1$};
    \draw[] (-1.7,4) node{$r_0$};
    \draw[] (1,3.5) node{$\bar r$};
\draw [] (1,6.2) node{$Q_b$};
    \draw[] (-1.7,1.1) node{$R$};
    \draw[black,thick] (-4,-4) rectangle (5,8);
  \end{scope}
    \begin{scope}[canvas is xz plane at y=3]
    \draw[dashed] (0,0) -- (0,6);
    \draw [] (0,6.4) node{$Q$};
    \draw [] (0,-0.3) node{$\bar z$};
    \filldraw [black] (0,6) circle (2pt);
  \end{scope}
      \begin{scope}[canvas is xy plane at z=6]
\draw[dashed] (0,0)-- (0,3);
  \end{scope}
    \draw[thick,-Latex] (0,0,0) -- (xyz cylindrical cs:radius=4);
  \draw[thick,-Latex] (0,0,0) -- (xyz cylindrical cs:radius=4,angle=90);
   \draw [thick,-Latex](0,0,6.1) -- (xyz cylindrical cs:z=7);
   \draw[] (0,0,-1.5) node {$A$};
\end{tikzpicture}
    \caption{The grey area represents a spherical region $A$ with radius $R$, centered at the origin. The region $A$ is layered with partial spherical shells. These shells are centered at the boundary-projected point $(\bar r,z=0,\phi_i=0)$ with a radius $r_0>\bar r-R$.
    }
    \label{fig:high-d-sphe}
\end{figure}

Now we turn to the \emph{bit thread flow} $V_{A}$. For example, given a spherical region $A=\{\textbf x||\textbf x|<R\}$, let us again confine the analysis on the $\phi_i=0$ slice and first consider the cases with  $\bar r>R$. In other words, the projection point $Q_b=(\bar r,0)$ of the bulk point $Q=(\bar r,\bar z)$ on the boundary is outside $A$.
Our strategy is that, we first decompose the region $A$ into layers of partial spherical shells $\bigodot_{r_0}$ centered at the projection point $Q_b$, with radius $r_0>\bar r-R$ (see Fig.\ref{fig:high-d-sphe}). The angular coordinate that parameterize the shell is $\phi_1$. Then we use \eqref{vx=0-HD} to calculate the contribution to $V_{A}(Q)$ from each shell. And finally, we integrate the contribution from all these shells to get $V_A(Q)$.

The area of a partial spherical shells $\bigodot_{r_0}$ is given by,
\begin{equation}\label{eq:area_shell_less_pi/2}
\begin{aligned}
        \int_0^{\bar\phi_{1}} \D \phi_1 \Omega_{d-3}r_0^{d-2}\sin^{d-3}\phi_1,
\end{aligned}
\end{equation}
where the upper limit of the integration $\bar\phi_{1}$ is given by,
\begin{equation}
    \cos(\bar\phi_{1})=\frac{r_0^2+\bar r^2-R^2}{2r_0\bar r}.
\end{equation}
Since $V_A$ is $\phi_i$-rotational symmetrical, the direction of $V_A(Q)$ lies exactly at the $\phi_i=0$ slice, which means we only need to compute $V^r_A(Q)$ and $V^z_A(Q)$.  The reason we classify the points using the such spherical shells is that, points in the same shell has the same distance from the projection point $Q_b$, hence their contribution to $V_A(Q)$ takes the same formula as \eqref{vx=0-HD}, with $\bar{r}$ replaced by their distance to $Q_b$ and $r$ representing the direction pointing to $Q_b$. Then it is straightforward to write down the contribution to $V^z_A(Q)$ from  $\bigodot_{r_0}$, which is just the summation from all the points on $\bigodot_{r_0}$,
\begin{equation}\label{vz}
\begin{split}
V^z_{\text{partial}\bigodot_{r_0}}(Q)
=&
\int_0^{\bar\phi_{1}} \D\phi_1
    V_{\textbf x=0}^z(r_0)\Omega_{d-3}r_0^{d-2} \sin^{d-3}\phi_1\\
    =&V_{\textbf x=0}^z(r_0)\Omega_{d-3}r_0^{d-2}\left(
    -_2F_1\left(\frac12,\frac{4-d}{2},\frac32,\cos^2\bar\phi_{1}\right)\cos\bar\phi_{1}+\frac{\pi^{3/2}\cos^{-1}(\frac{(d-3)\pi}{2})}{(d-3)\Gamma(\frac{4-d}{2})\Gamma(\frac{d-3}{2})}
    \right),
    \end{split}
\end{equation}
while the contribution to $V^r_A(Q)$ should have an additional multiplier $\cos\phi_1$ to get the projection on the $r$ direction,
\begin{equation}\label{vr}
\begin{split}
    V^r_{\text{partial}\bigodot_{r_0}}(Q)=&
   \int_0^{\bar\phi_{1}} \D\phi_1 V_{\textbf x=0}^r(r_0)\Omega_{d-3}r_0^{d-2} \sin^{d-3}\phi_1\cos\phi_1\\
    =&V_{\textbf x=0}^r(r_0)\Omega_{d-3}r_0^{d-2}\frac{\sin^{d-2}\bar\phi_{1}}{d-2}.
    \end{split}
\end{equation}
At last, we integrate over all the partial spherical shells to get the \emph{bit thread flow},
\begin{align}\label{eq:high-d,VAr}
 V_A^r(Q)=&\int_{\bar r-R}^{\bar r+R}\D r_0~ V^r_{\text{partial}\bigodot_{r_0}}
        =\frac{1}{4G_N}\frac{\bar r\bar z}{R}\left(
        \frac{2R \bar  z}{\sqrt{(R^2+\bar r^2+\bar z^2)^2-4R^2\bar r^2}}
        \right)^d,
        \\\label{eq:high-d,VAz}
V_A^z(Q)=&\int_{\bar r-R}^{\bar r+R} \D r_0~  V^z_{\text{partial}\bigodot_{r_0}} 
    =\frac{1}{4G_N}\frac{R^2-\bar r^2+\bar z^2}{2R}\left(
        \frac{2R \bar z}{\sqrt{(R^2+\bar r^2+\bar z^2)^2-4R^2\bar r^2}}
        \right)^d.
\end{align}

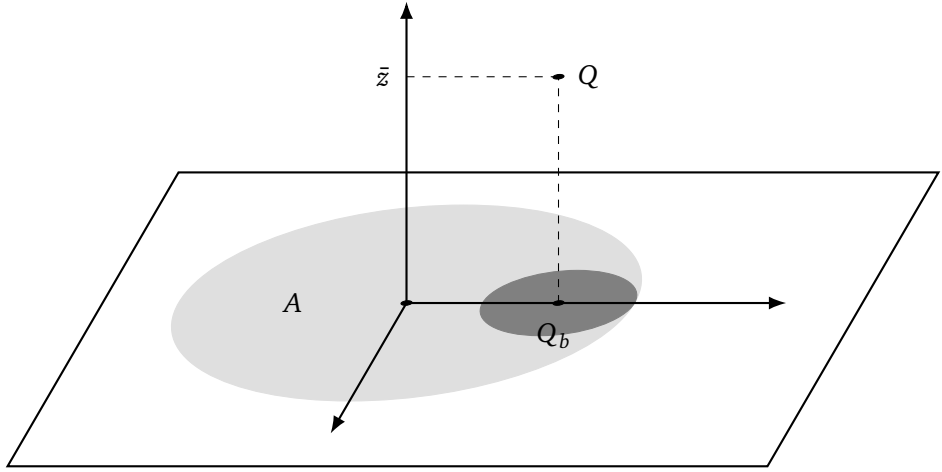
\begin{figure}
    \centering
    \begin{tikzpicture}[z={(0:10mm)},x={(-120:5mm)}]]
    \begin{scope}[canvas is xz plane at y=0]
        \filldraw[lightgray!50] (0,0) circle [radius=3];
    \filldraw[black] (0,0) circle (2pt);
    \filldraw[gray] (0,2) circle (1);
    \filldraw [black] (0,2) circle (2pt);
    \draw[] (1,2.2) node{$Q_b$};
    \draw[black,thick] (-4,-4) rectangle (5,6);
  \end{scope}
    \begin{scope}[canvas is xz plane at y=3]
    \draw[dashed] (0,0) -- (0,2);
    \draw [] (0,2.4) node{$Q$};
    \draw [] (0,-0.3) node{$\bar z$};
    \filldraw [black] (0,2) circle (2pt);
  \end{scope}
      \begin{scope}[canvas is xy plane at z=2]
\draw[dashed] (0,0)-- (0,3);
  \end{scope}
    \draw[thick,-Latex] (0,0,0) -- (xyz cylindrical cs:radius=4);
  \draw[thick,-Latex] (0,0,0) -- (xyz cylindrical cs:radius=4,angle=90);
   \draw [thick,-Latex](0,0,0) -- (xyz cylindrical cs:z=5);
   \draw[] (0,0,-1.5) node {$A$};
\end{tikzpicture}
    \caption{ The bulk point is projected inside the region $A$. The shaded area has no contribution to $V_A^r$.
    }
    \label{fig:high-d-sphe-2}
\end{figure}

Then we consider the case with $\bar r<R$, where the projection point $Q_b$ of $Q$ lies inside $A$. Following the same strategy as the case with $\bar r>R$, the region $A$ is also foliated by layers of partial spherical shells centered at $\bar r$ and one may obtain $V^\mu_A$ by integrating over all the shells. 
The only difference is that, in the spherical region centered at $Q_b$ with radius $R-\bar r$, the whole spherical shell contribute to $V_A$ (see Fig.\ref{fig:high-d-sphe-2}).  For symmetry reason, this spherical region has no contribution to $V^r_A$, and its contribution to $V^z_A$ is computed by,
\begin{equation}
\begin{split}
\int _0^{R-\bar r}\D r_0V^z_{\text{whole}\bigodot_{r_0}}
=&
    \int_0^{R-\bar r}\D r_0 \int_0^{\pi} \D\phi_1  V_{\textbf x=0}^z(r_0)\Omega_{d-3}r_0^{d-2}\sin^{d-3}\phi_1
    \end{split}.
\end{equation}
The contribution from the other points on the partial spherical shell can be carried out following our discussion for the case with $\bar{r}>R$. Then we get the \emph{bit thread flow} $V_A(Q)$
\begin{equation}
    \begin{aligned}
        V_A^r(Q)=&\int_{R-\bar r}^{\bar r+R}\D r_0~ V^r_{\text{partial}\bigodot_{r_0}}
        =\frac{1}{4G_N}\frac{\bar r\bar z}{R}\left(
        \frac{2R \bar  z}{\sqrt{(R^2+\bar r^2+\bar z^2)^2-4R^2\bar r^2}}
        \right)^d,
    \end{aligned}
\end{equation}
and 
\begin{equation}
\begin{aligned}
       V_A^z(Q)
        =&\int _0^{R-\bar r}\D r_0V^z_{\text{whole}\bigodot_{r_0}}+\int_{R-\bar r}^{\bar r+R} \D r_0~  V^z_{\text{partial}\bigodot_{r_0}} \\
    =&\frac{1}{4G_N}\frac{R^2-\bar r^2+\bar z^2}{2R}\left(
        \frac{2R \bar z}{\sqrt{(R^2+\bar r^2+\bar z^2)^2-4R^2\bar r^2}}
        \right)^d.
\end{aligned}
\end{equation}
The above results share the same expressions as \eqref{eq:high-d,VAr} and \eqref{eq:high-d,VAz}, which means we do not need to distinguish between the case with $\bar{r}>R$ and the case with $\bar{r}>R$. 

Remarkably, in general dimensions the resulting $V^\mu_A$ also coincides with the bit threads configuration \eqref{bt-agon} constructed in \cite{Agon:2018lwq}.

\section{PEE threads for multi-intervals}\label{sec:pt-pt}

\begin{figure}
    \centering
     \begin{tikzpicture}[scale=1]
\clip (-6,-1.5) rectangle (6,5);
\filldraw[LouisBlue!30](-4,0) arc (180:0:4);
\filldraw[LouisBlue!30](1,0) arc (180:0:1);
\filldraw[white](-1,0) arc (180:0:1);
\draw[ultra thick,LouisBlue] (-4,0) arc (180:0:4) ;
\draw[ultra thick,LouisBlue] (-1,0) arc (180:0:1) ;
\draw[very thick] (-9,0) -- (9,0);
\draw[] (0,-0.4) node {$B_1$};
\draw[] (2.5,-0.4) node {$A_2$};
\draw[] (-2.5,-0.4) node {$A_1$};
\draw[] (-5,-0.4) node {$B_2$};
\draw[] (5,-0.4) node {$B_2$};
\draw[ thick,dashed,LouisColor1] (-2,0) arc (180:0:2) ;
\draw[ thick,dashed,LouisOrange] (0,0) arc (180:0:2.5) ;
\draw[ thick,dashed,LouisColor2] (-3,0) arc (0:180:1) ;
\draw[ thick,dashed,LouisOrange] (-5.7,3) -- (-5,3) ;
\draw[] (-4.3,3.05) node {$\omega=2$};
\draw[ thick,dashed,LouisColor1] (-5.7,2.) -- (-5,2.) ;
\draw[] (-4.3,2.05) node {$\omega=0$};
\draw[thick,dashed,LouisColor2] (-5.7,2.5) -- (-5,2.5) ;
\draw[] (-4.3,2.55) node {$\omega=1$};
\draw[] (0,4.5) node {$\alpha>1/2$};
\end{tikzpicture}
    \begin{tikzpicture}[scale=1]
\clip (-6,-1.5) rectangle (6,4);
\filldraw[LouisBlue!30](-4,0) arc (180:0:1.5);
\filldraw[LouisBlue!30](1,0) arc (180:0:1.5);
\draw[ultra thick,LouisBlue] (-4,0) arc (180:0:1.5) ;
\draw[ultra thick,LouisBlue] (1,0) arc (180:0:1.5) ;
\draw[very thick] (-9,0) -- (9,0);
\draw[] (0,-0.4) node {$B_1$};
\draw[] (2.5,-0.4) node {$A_2$};
\draw[] (-2.5,-0.4) node {$A_1$};
\draw[] (-5,-0.4) node {$B_2$};
\draw[] (5,-0.4) node {$B_2$};
\draw[ thick,dashed,LouisColor1] (-1.5,0) arc (180:0:1.5) ;
\draw[ thick,dashed,LouisOrange] (0,0) arc (180:0:2.5) ;
\draw[ thick,dashed,LouisColor2] (-5,0) arc (180:0:1) ;
\draw[ thick,dashed,LouisColor1] (-5.7,3) -- (-5,3) ;
\draw[] (-4.3,3.05) node {$\omega=2$};
\draw[ thick,dashed,LouisOrange] (-5.7,2.) -- (-5,2.) ;
\draw[] (-4.3,2.05) node {$\omega=0$};
\draw[thick,dashed,LouisColor2] (-5,0) arc (180:0:1) ;
\draw[thick,dashed,LouisColor2] (-5.7,2.5) -- (-5,2.5) ;
\draw[] (-4.3,2.55) node {$\omega=1$};
\draw[] (0,3.7) node {$0<\alpha<1/2$};
\end{tikzpicture}
    \caption{Connected phase (the upper figure) and disconnected phase (the lower figure) for $A=A_1\cup A_2$. 
The threads connecting $A_1$ and $A_2$ in disconnected phase are \emph{outer threads} as they pass through the boundary of $\mathcal W_A$ (blue shaded regions), while those in connected phase are \emph{inner threads} as they are confined in $\mathcal W_A$.
    }
    \label{fig:pt-pt}
\end{figure}
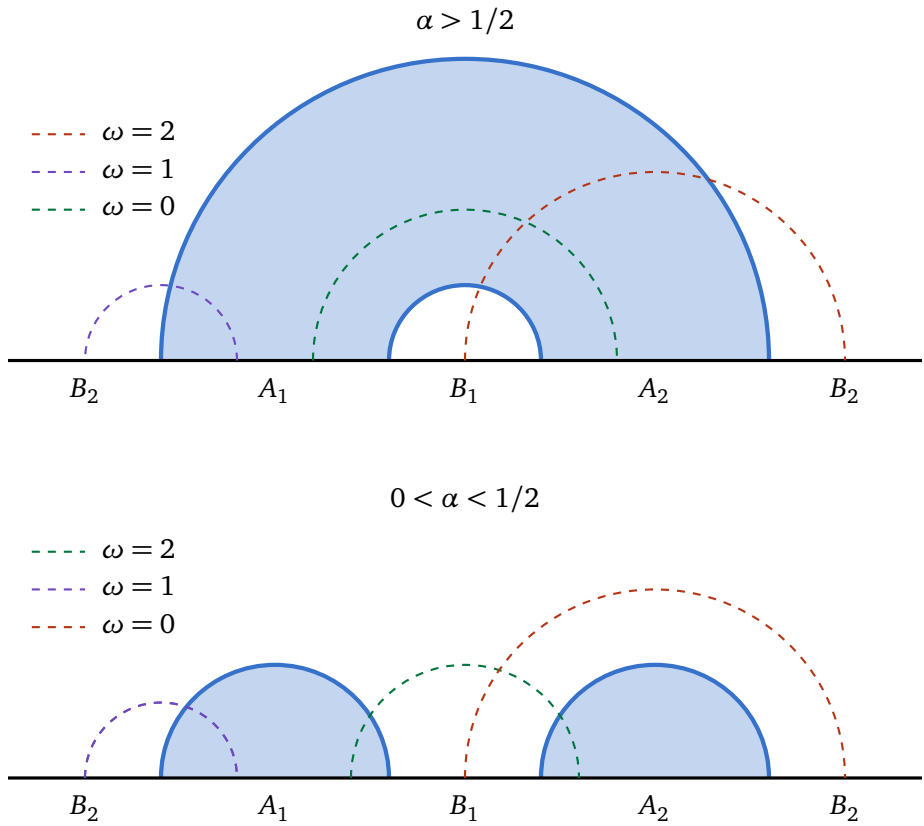

Previously, we mainly focus on the PEE thread configurations for intervals or spherical boundary regions, which are connected regions. We have shown that the entanglement entropy is given by the flux associated with all the \emph{outer threads} of the connected region. In this section, we turn to the cases of disconnected intervals and show how the holographic entanglement entropy can be reproduced by counting certain classes of PEE threads in the bulk. This is a non-trivial problem as the entanglement entropy for disconnected regions undergoes a phase transition where the RT surfaces in the bulk changes discontinuously \cite{Hartman:2013mia}. Upon such phase transitions, the bit thread configurations should also change while the PEE thread configurations should not change. A naive generalization of the scheme in the previous section cannot reproduce the holographic entanglement entropy for disconnected regions \cite{Wen:2019iyq}. In the following, we propose that due to the phase transition of the entanglement wedge, new rules to define the \emph{inner} and \emph{outer PEE threads} should be taken into account to reproduce the holographic entanglement entropy.

In this section we focus on the case of Poincar\'e AdS$_3$. Consider a disconnected interval $A=A_1\cup A_2$ and its complement $B=B_1\cup B_2$ on the boundary, and $A\cup B$ makes up the vacuum state of the boundary holographic CFT$_2$.  
Let us mark $A_1=[a_1,a_2]$ and $A_2=[a_3,a_4]$, and mark the entanglement wedge $\mathcal{W}_{A}$ of $A$ in blue (see Fig.\ref{fig:pt-pt}). According to the RT formula \cite{Ryu:2006bv,Ryu:2006ef,Hartman:2013mia} when the cross ratio 
\begin{align}
\alpha=\frac{(a_2-a_1)(a_4-a_3)}{(a_3-a_1)(a_4-a_2)}>\frac{1}{2},
\end{align}
$\mathcal{W}_A$ is connected and we have $\mathcal{W}_A\supset\mathcal{W}_{A_1}\cup\mathcal{W}_{A_2}$. In this connected phase the entanglement entropy is calculated by
\begin{equation}
 \textit{Connected phase}:  \quad  S_{A}=S_{B_1}+S_{A_1 B_1 A_2},
\end{equation}
where $B_1$ is the sandwiched interval between $A_1$ and $A_2$. As $A_1$ and $A_2$ get far apart $0<\alpha<\frac{1}{2}$, the $\mathcal{W}_{A}$ becomes disconnected and $\mathcal{W}_{A}=\mathcal{W}_{A_1}\cup\mathcal{W}_{A_2}$. In this disconnected phase, the entanglement entropy becomes
\begin{equation}\label{SEEdis}
\textit{Disconnected phase}: \quad    S_A=S_{A_1}+S_{A_2} .
\end{equation}

Previously we define the PEE threads that connecting one point inside $A$ and another point outside $A$ as the \emph{outer threads}, which are the threads that contribute non-trivially to the entanglement entropy. A simple example to illustrate this problem is the two interval in disconnected phase. If we stick to this definition for the inner and \emph{outer threads}, then the PEE threads connecting points in $A_1$ and points in $A_2$ should be \emph{inner threads} hence do not contribute to $S_A$, and we get\begin{align}
 S_{A}=\mathcal{I}(A,B)=\mathcal{I}(A_1,B)+\mathcal{I}(A_2,B)\,.
 \end{align} 
This is not consistent with the RT formula \eqref{SEEdis}. Also, it was claimed in \cite{Wen:2019iyq} that, the normalization property $S_{A}=\mathcal{I}(A,B)|_{B\to\bar{A}}$ of the PEE does not hold for the disjoint intervals. On the right hand side of \eqref{SEEdis}, $S_{A_1}$ contains all the \emph{outer threads} of $A_1$, which can be decomposed into $\mathcal{I}(A_1,B)\cup \mathcal{I}(A_1,A_2)$, and a similar decomposition applies to $S_{A_2}$ hence we should have
\begin{align}
S_{A}=\mathcal{I}(A_1,B)+\mathcal{I}(A_2,B)+2\mathcal{I}(A_1,A_2)\,.
\end{align}
The above equation indicates that, the PEE threads connecting $A_1$ and $A_2$ not only contribute non-trivially to $S_{A}$, but also to $S_{B}$ as $S_{A}=S_{B}$. More interestingly, the contribution from such PEE threads is doubly counted. 

For the purpose to reproduce the holographic entanglement entropy from the PEE threads, we should reconsider the classification of \emph{inner} and \emph{outer threads}. Also, note that the PEE threads connecting $A_1$ and $A_2$ in the disconnected phase pass through the entanglement wedge $\mathcal{W}_{B}$ twice in the bulk. This is a new phenomenon compared with the configurations where $A$ is a single interval and the \emph{inner threads} are all confined in $\mathcal{W}_{A}$. The holographic entanglement entropy between $A$ and $B$ is indeed the generalized gravitational entanglement entropy between $\mathcal{W}_{A}$ and $\mathcal{W}_{B}$ \cite{Lewkowycz:2013nqa}. If the PEE threads are some physical objects in the bulk that representing the entanglement flow in the dual gravity, then it is natural to count the threads that pass through the boundaries between $\mathcal{W}_{A}$ and $\mathcal{W}_{B}$. 

Inspired by the above discussions, we give a new definition for the \emph{inner} and \emph{outer PEE threads} based on the entanglement wedge configuration, which solves all the problems one has in AdS$_3$. Given two (connected or disconnected) regions $A$ and $B$, which together make up the whole boundary, and their entanglement wedges in the bulk, the \emph{inner} and \emph{outer threads} are defined in the following:
\begin{itemize}
    \item 
 \textit{ outer threads}:\quad PEE threads that pass through the boundary between $\mathcal{W}_{A}$ and $\mathcal{W}_{B}$.
  \item 
  \textit{ inner threads}:\quad PEE threads that are confined inside $\mathcal{W}_{A}$ or $\mathcal{W}_{B}$.
\end{itemize}
In addition we also define a new parameter for the \emph{outer threads},
\begin{itemize}
\item \textit{weight of an outer thread} $\omega$: the number of the times that the \textit{outer thread} passes through the boundary between $\mathcal{W}_{A}$ and $\mathcal{W}_{B}$.
The \emph{inner threads} are just threads with $\omega=0$.
\end{itemize}
For any connected regions $A$ and $B$, we also refer to $\omega_{AB}=\omega_{BA}$ as the weight of the threads connecting $A$ and $B$. The threads with $\omega=0$ are just \emph{inner threads}. Obviously, the above definition reduces to the definition in section \ref{sec:geo-sch} for the cases of $A$ being single intervals.

Let us use the above definitions to explain the entanglement entropy $S_{A}$ in the connected phase (see the upper figure in Fig.\ref{fig:pt-pt}). In this case, given $i,j=1,2$ we have $\omega_{A_i A_j}=0$, $\omega_{A_i B_j}=1$ and $\omega_{B_i B_j}=2$, thus, by counting the PEE threads with different weights, the entanglement entropy is given by
\begin{align}
S_{A}=&\mathcal{I}(A_1,B_1\cup B_2)+\mathcal{I}(A_2,B_1\cup B_2)+2\mathcal{I}(B_1,B_2)
\cr
=&\mathcal{I}(A_1\cup A_2\cup B_2,B_1)+\mathcal{I}(A_1\cup A_2\cup B_1,B_2)
\cr
=&S_{B_1}+S_{B_2}\,,
\end{align}
which coincides with the RT formula.

\begin{figure}
\centering
\begin{tikzpicture}[scale=1]
\clip (-10,-1.5) rectangle (5,5);
\filldraw[LouisBlue!30](-3,0) arc (180:0:3);
\filldraw[LouisBlue!30](1,0) arc (180:0:1);
\filldraw[white](-1,0) arc (180:0:1);
\filldraw[LouisBlue!30](-8,0) arc (180:0:1);
\draw[ultra thick,LouisBlue] (-3,0) arc (180:0:3) ;
\draw[ultra thick,LouisBlue] (-1,0) arc (180:0:1) ;
\draw[ultra thick,LouisBlue] (-8,0) arc (180:0:1) ;
\draw[very thick] (-12,0) -- (9,0);
\draw[] (0,-0.4) node {$B_2$};
\draw[] (2,-0.4) node {$A_3$};
\draw[] (-2,-0.4) node {$A_2$};
\draw[] (-4.5,-0.4) node {$B_1$};
\draw[] (4.5,-0.4) node {$B_3$};
\draw[] (-9,-0.4) node {$B_3$};
\draw[] (-7,-0.4) node {$A_1$};
\draw[ thick,dashed,LouisColor1] (-2,0) arc (180:0:2) ;
\draw[ thick,dashed,LouisOrange,inversemiddlearrow={0.17},inversemiddlearrow={0.11},middlearrow={0.4},middlearrow={0.45}] (0,0) arc (180:0:2.5) ;
\draw[ thick,dashed,LouisColor2, middlearrow={0.35},middlearrow={0.5}] (-2.2,0) arc (0:180:1);
\draw[ thick,dashed,black,middlearrow={0.08},middlearrow={0.13},inversemiddlearrow={0.7},inversemiddlearrow={0.75},middlearrow={0.9},middlearrow={0.95},] (-7,0) arc (180:0:3.4) ;
\draw[ thick,dashed,LouisColor2] (-10,2.5) -- (-9,2.5) ;
\draw[ thick,dashed,LouisOrange] (-10,3) -- (-9,3) ;
\draw[ thick,dashed,LouisColor1] (-10,2.) -- (-9,2.) ;
\draw[ thick,dashed,black] (-10,3.5) -- (-9,3.5) ;
\draw[] (-8.3,2.55) node {$\omega=1$};
\draw[] (-8.3,3.05) node {$\omega=2$};
\draw[] (-8.3,2.05) node {$\omega=0$};
\draw[] (-8.3,3.55) node {$\omega=3$};
\end{tikzpicture}
\caption{The entanglement wedge and PEE threads with different weights for $A=A_1\cup A_2\cup A_3$ with disconnected $\mathcal W_{A_1\cup A_2\cup A_3}$ and connected $\mathcal W_{A_2\cup A_3}$. Here the direction of the PEE flow is from the inside to the outside of the entanglement wedge.}
\label{3-intervals}
\end{figure}
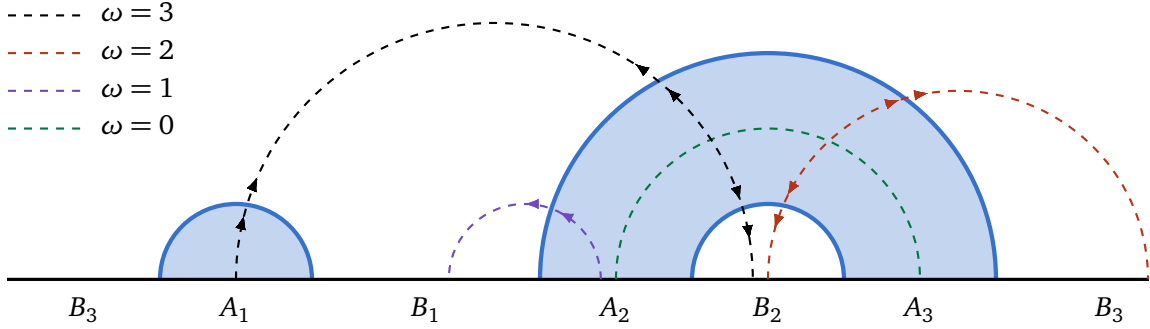

We can further explore more complicated scenarios, such as a multi-interval configuration $A=A_1\cup A_2\cup A_3$ depicted in Fig.\ref{3-intervals}. In this setup, $A_1$ is situated sufficiently far from $A_2$ and $A_3$ so that the entanglement wedge $\mathcal W_{A_1\cup A_2\cup A_3}$ becomes disconnected. Moreover, $A_2$ and $A_3$ maintain close proximity, resulting in a connected $\mathcal W_{A_2\cup A_3}$.
The complementary region to $A$ is identified as $B=B_1\cup B_2\cup B_3$. Here, $B_2$ is settled between $A_2$ and $A_3$, while $B_1$ is flanked by $A_1$ and $A_2$.
 According to the RT formula, the entanglement entropy is given by:
\begin{align}
S_{A}
=&S_{A_1}+S_{B_2}+S_{A_2\cup B_2\cup A_3}\,.
\end{align}
To recover the RT result, we read out the weights of \emph{outer threads} from the entanglement wedge configuration Fig.\ref{3-intervals}.
The weights of the \emph{outer threads} are presented in Table \ref{tab:1}. By counting the \emph{outer threads}, we obtain the entanglement entropy:
\begin{align}
S_{A}=&\mathcal{I}(A_1,B_1 B_3)+2\mathcal{I}(A_1,A_2A_3)+3\mathcal{I}(A_1,B_2)+\mathcal{I}(A_2A_3,B_1 B_2B_3)+2\mathcal{I}(B_1B_3,B_2)
\notag\\
=&\mathcal{I}(A_1,A_2A_3B_1B_2 B_3)+\mathcal{I}(B_2,A_1A_2A_3B_1 B_3)+\mathcal{I}(A_2 B_2 A_3,A_1B_1B_3)
\notag\\
=&S_{A_1}+S_{B_2}+S_{A_2\cup B_2\cup A_3},
\end{align}
This result aligns with the RT formula.
\begin{table*}[htp]
  \centering
  \renewcommand\tabcolsep{1.8pt}
  \begin{tabular}{ccccccc}
  \hline
   $\omega$ &$A_1$&$A_2$ & $A_3$ & $B_1$ & $B_2$ & $B_3$\\
  \hline
  $A_1$& 0 & 2 & 2  & 1 & 3 & 1\\
  $A_2$& 2 &0 & 0  & 1& 1 & 1 \\
  $A_3$& 2 &0 & 0  & 1 & 1 & 1\\
  $B_1$&1 &1& 1  & 0 & 2 & 0 \\
  $B_2$ & 3  &1 & 1  & 2 & 0 & 2\\
  $B_3$ & 1  &1 & 1  & 0 & 2 & 0 \\
  \hline
  \end{tabular}
 \caption{The weight of PEE threads for $A=A_1\cup A_2\cup A_3$ with disconnected $\mathcal W_{A_1\cup A_2\cup A_3}$ and connected $\mathcal W_{A_2\cup A_3}$. }
\label{tab:1}
\end{table*}

Note that the above definition for the \emph{inner} and \emph{outer threads} and the weight of the threads depends on an explicit configuration for the entanglement wedge. One may conclude that, the PEE threads may not be a complete reformulation of the RT formula in AdS$_3$. Nevertheless, we can get rid of this dependence by considering all the possible assignments for the weights and choose the one that gives the minimal value for the entanglement entropy $S_{A}$.

More explicitly, for a region $A$ with multi sub-intervals $A=\cup A_i$ and its complement $B=\cup B_{j}$, let us consider an arbitrary surface $\Sigma_{A}$ in the bulk that is homologous to $A$, thus we can get a corresponding assignment for the weights $\omega_{A_iA_j}$, $\omega_{A_i B_j}$ and $\omega_{B_i B_j}$ by counting the number of times the thread intersect with the surface. Then the entanglement entropy is given by minimizing the following summation,
\begin{align}\label{RTPEE}
    S_{A}=S_{B}=\text{min}_{\Sigma_{A}}\left[\sum_{i,j}\omega_{A_iA_j}\mathcal{I}(A_i,A_j)+\sum_{i,j}\omega_{B_iB_j}\mathcal{I}(B_i,B_j)+\sum_{i,j}\omega_{A_iB_j}\mathcal{I}(A_i,B_j)\right]
    \end{align}
The above proposal modifies the naive normalization requirement $S_{A}=\mathcal{I}(A,\bar{A})$. For any $\Sigma_{A}$, the above summation actually calculate the flux of the PEE threads passing through $\Sigma_{A}$. The homologous surface that minimizes the above summation will coincide with the RT surface. In the following, we will give simple examples to test this proposal. See \cite{Lin:2024dho} for a general proof for this statement. Since $A$ and $B$ are separated by a homologous surface, we should in general have that:
\begin{itemize}
\item $\omega_{A_iB_j}$ should be odd numbers, i.e. $\omega_{A_iB_j}=1,3,5\cdots$,

\item $\omega_{A_iA_j}$ and $\omega_{B_iB_j}$ can be zero or non-zero even numbers, i.e. $\omega_{A_iA_j},\omega_{B_iB_j}=0,2,4\cdots$.
\end{itemize}

Now we revisit the two interval case shown in Fig.\ref{fig:pt-pt}. 
In this case the minimal weight of $\omega_{A_iB_j}$ is 1. While the minimal weight for $\omega_{A_1A_2}$ and $\omega_{B_1B_2}$ is zero. Nevertheless, they can not be zero simultaneously, since there is no homologous surface that is consistent with the assignment $\omega_{A_1A_2}=\omega_{B_1B_2}=0$. So we conclude that, the possible assignments for the weights that may give the minimal $S_{A}$ are listed in the following:
\begin{enumerate}
    \item $\omega_{\omega_{A_iB_j}}=1$,\ $\omega_{A_1A_2}=0$,\ $\omega_{B_1B_2}=2$,
     \item $\omega_{\omega_{A_iB_j}}=1$,\ $\omega_{A_1A_2}=2$,\ $\omega_{B_1B_2}=0$.
\end{enumerate}
Then the corresponding entanglement entropy $S_{A}$ is calculated by \eqref{RTPEE}, which is the minimal weighted summation between the following two results:
\begin{enumerate}
    \item 
$\mathcal{I}(A,B)+2\mathcal{I}(B_1,B_2)
=S_{A_1B_2A_2}+S_{B_2}$.
     \item 
$\mathcal{I}(A,B)+2\mathcal{I}(A_1,A_2)=S_{A_1}+S_{A_2}$.
\end{enumerate}
When $A_1$ and $A_2$ is far enough such that $0<\alpha<1/2$, the first result will give the minimal value for $S_{A}$. While the second result minimizes $S_{A}$ when $A_1$ and $A_2$ get closer such that $\alpha>1/2$. This gives a complete reformulation for the phase transition of the RT surface in terms of the PEE threads.

\section{Discussions}\label{sec:con-dis}
In this paper, we propose a natural scheme to geometrize the PEE using bulk geodesics, which we refer to as PEE threads. The configuration of these PEE threads is solely determined by the state and represents an intrinsic structure that is independent of the specific region under consideration. In the context of Poincaré AdS spacetime, we demonstrate that for any static interval or spherical region, a unique configuration of bit threads can emerge from the PEE thread configuration by superimposing all the \emph{PEE thread flows} emanating from that region. These bit thread configurations are regarded as the most natural ones as they respect the symmetries of the system and possess a clear physical interpretation inherited from PEE.

In AdS$_3$/CFT$_2$, we further study the reformulation of the RT formula for multi-interval regions based on the PEE thread configuration. We find that, the PEE threads should be weighted by the number of times that they intersect with the RT surface. Remarkably, we can get rid of the predetermined RT surface by considering all the possible assignments that are consistent with any homologous surface of the region, and choosing the assignment that gives the minimal summation \eqref{RTPEE}. The assignment that minimizes the summation will correspond to certain type of homologous surface, which is exactly the RT surface if we require it to be extremal. This reformulation also works for single intervals. In summary, based on the PEE thread configuration we gave a reformulation of the RT formula for general static regions in the vacuum state of AdS$_3$/CFT$_2$, and for static spherical regions in higher dimensions.

For the cases of multi intervals, since the PEE threads can intersect with the RT surface multiple times, the claim that there is a definite direction for any PEE threads does not make sense. This means our strategy to generate bit thread configurations from PEE threads breakdown. Nevertheless, we think the formulation of weighted PEE threads makes more sense. The reason is that, the holographic entanglement entropy is the generalized gravitational (entanglement) entropy \cite{Lewkowycz:2013nqa} between entanglement wedges $\mathcal{W}_{A}$ and $\mathcal{W}_{B}$, which includes different bulk regions at the critical point. It is reasonable that, the parts of the PEE threads in different entanglement wedges play different roles. One can also define an in-definite flow direction which is always locally from one side of the homologous surface to the other side (see the arrows in Fig. \ref{3-intervals}). Then the norm of the PEE flow on homologous surface can recover $1/(4G)$ only if this homologous surface coincide with the RT surface.

The success of our reformulation indicates that the PEE threads may be a certain physical quantity that lies in the bulk gravity theory. It seems that, the part of the PEE thread that lies in $\mathcal{W}_{A}$ should be considered as degrees of freedom in $\mathcal{W}_{A}$ while the other part belongs to $\mathcal{W}_{B}$. The entanglement entropy between the two parts at leading order is just proportional to the number of partition points, i.e. the weight $\omega$. It will be very interesting to explore a deeper physical interpretation of the PEE threads in the bulk (quantum) gravity or its toy models. For example, the PEE threads is very reminiscent of the so-called pentagon-edge geodesics defined in the toy model of AdS$_3$/CFT$_2$ based on fracton models with subsystem symmetry \cite{Yan:2018nco,Yan:2019quy}. More interestingly, the network of the PEE threads could be considered as a well-defined continue limit of a bulk tensor network that describes the multi-scale entanglement structure of the holographic CFT. See \cite{Swingle:2009bg,Hayden:2016cfa,Pastawski:2015qua} for some earlier toy models along this line.

One may wonder if we can generate a bit thread configuration for non-spherical but connected regions in higher dimensions. We give a simple exploration for the \emph{bit thread flow} $V_A$ for a strip in Appendix.\ref{app:strip}. Unfortunately, as we can see $|V_A|<1/4G_N$ on the RT surface of strip and thus does not satisfy the requirements for bit threads. Moreover, a naive integration of the PEE flow does not reproduce the entanglement entropy calculated by the RT formula. One important reason for this failure could be the fact that, unlike the spherical regions, the modular Hamiltonian is non-local for strip. Another possible reason is that, unlike the spherical regions, the PEE threads with both its endpoints inside or outside the region can also intersect with the RT surface twice. This indicates that we should classify the PEE threads and weight them differently as in the multi-interval cases, hence the entanglement entropy may coincide with the RT formula.  Indeed this is a non-trivial task even for strips. We leave this for further investigations.

There are absolutely many other interesting future directions relevant to the PEE threads. For example, we can explore the problems of how to generalize the PEE threads to its covariant version, how to extend our discussion from Poincar\'e AdS to AdS black holes and how to define a quantum version of PEE threads to include the quantum correction \cite{Faulkner:2013ana} in the bulk gravity as was done in \cite{Rolph:2021hgz,Agon:2021tia} for a quantum version of bit threads. The PEE has been explored in holographic theories beyond AdS/CFT \cite{Wen:2018mev,Wen:2020ech,Camargo:2022mme,Basu:2022nyl,Liu:2023djf}. Generalizing the PEE threads to more generic holographic theories is also very interesting.

\textit{Note added}: Two months after this paper was submitted to arxiv, a new submission \cite{Lin:2024dho} by the same authors appears on arxiv. In the follow-up paper we proved that, in the vacuum state of AdS/CFT, for any static boundary region $A$, including those non-spherical regions and disconnected regions, the homologous surface $\Sigma_{A}$ that has the minimal flux of the PEE threads passing through it is exactly the Ryu-Takayanagi (RT) surface of $A$, and the minimal flux coincides with the holographic entanglement entropy of $A$. Furthermore, we can extract much more information from the configuration of the PEE threads than the RT formula. We showed that the strength of the PEE flux at any bulk point along any direction is $1/4G$, which means the AdS space is full of PEE threads. Based on this observation, we proved that any co-dimension two surfaces in the bulk can be reconstructed by the PEE threads passing through it.

\section*{Acknowledgment}
J. Lin is supported by the National Natural Science Foundation of China under Grant No.12247117, No.12247103 and No.12047502.
Y. Lu is supported by the National Natural Science Foundation of China under Grant No.12247161, the NSFC Research Fund for International Scientists (Grant No. 12250410250), and the China Postdoctoral Science Foundation under Grant No.2022TQ0140.

\appendix

\section{Proof of a statement}\label{app:pos}
\begin{figure}
    \centering
\begin{tikzpicture}[scale=0.6]
        \draw[LouisOrange,middlearrow={0.37},dashed,thick](-1,0) arc (180:0:2.59376);
    \draw[LouisOrange,middlearrow={0.31},dashed,thick](2,0) arc(0:180:2.83151);
    \draw[LouisBlue,ultra thick](2-2*2.83151,0) arc (180:0:2.59376-0.5-1+2.83151);
    \draw[very thick] (-4.4,0) -- (5,0);
    \draw[thick,LouisColor1] (-2.5,0) arc (180:0:2.5) ;
    \draw[gray,dashed,middlearrow={1},thick](0,0) -- (0.905867,3.38073) ;
    \draw[] (0.505867,2.8) node[label={[scale=0.6]$V_A^\mu$}]{};
\draw[] (-1,-0.75) node[label={[scale=0.6]$r_1$}]{};
\draw[] (2,-0.75) node[label={[scale=0.6]$r_2$}]{};
    \filldraw [black] (0.647048,2.41481) circle (2pt);
    \draw[](0.747048,1.5) node[label={[scale=0.6]$Q$}]{};
    \draw[LouisOrange,ultra thick](-1,0)--(2,0);
    \filldraw[black](-1,0) circle (2pt);
    \filldraw[black](2,0) circle (2pt);
    \filldraw[LouisColor1](-2.5,0) circle (2pt);
    \filldraw[LouisColor1](2.5,0) circle (2pt);
    \draw[](-2,1.6) node[label={[scale=0.6]{\color{LouisColor1}$\mathcal E_{R_Q}$}}]{};
\draw[](-2,3.2) node[label={[scale=0.6]$\mathcal E_{A}$}]{};
\end{tikzpicture}
\begin{tikzpicture}[scale=0.6]
        \draw[LouisOrange,middlearrow={0.37},dashed,thick](-1,0) arc (180:0:2.59376);
    \draw[LouisOrange,middlearrow={0.31},dashed,thick](2,0) arc(0:180:2.83151);
    \draw[LouisBlue,ultra thick](2-2*2.83151,0) arc (180:0:2.59376-0.5-1+2.83151);
    \draw[very thick] (-6,0) -- (5,0);
    \draw[ thick,LouisColor1] (0.966486+0.647048,0) arc (0:180:3.5) ;
    \draw[gray,dashed,middlearrow={1},thick](0.966486+0.647048-3.5,0) -- (1.4071,2.41481*1.3);
    \draw[] (0.505867,2.8) node[label={[scale=0.6]$V_A^\mu$}]{};
\draw[] (-1,-0.75) node[label={[scale=0.6]$r_1$}]{};
\draw[] (2,-0.75) node[label={[scale=0.6]$r_2$}]{};
    \filldraw [black] (0.647048,2.41481) circle (2pt);
    \draw[](0.7,1.45) node[label={[scale=0.6]$Q$}]{};
    \draw[LouisOrange,ultra thick](-1,0)--(2,0);
    \filldraw[black](-1,0) circle (2pt);
    \filldraw[black](2,0) circle (2pt);
    \filldraw[LouisColor1](0.966486+0.647048-7,0) circle (2pt);
    \filldraw[LouisColor1](2.5,0) circle (2pt);
    \filldraw[LouisColor1](2.5,0) circle (2pt);
    \draw[](-4,1.6) node[label={[scale=0.6]{\color{LouisColor1}$\mathcal E_{R_Q}$}}]{};
\draw[](2.2,3.2) node[label={[scale=0.6]$\mathcal E_{A}$}]{};
\end{tikzpicture}
    \caption{
    The total vector field $V^{\mu}_A$ at $Q$ receives the net contributions from the PEE threads of the red boundary region.
    For $Q$ inside the $\mathcal{W}_A$, this red region $[r_1,r_2]$ is determined by two intersecting geodesics (red dashed curves).
    The green dashed semicircle denotes the RT surface $\mathcal E_{R_Q}$ passing through the field point $Q$ and normal to $V^{\mu}_A$. $R_Q$ is the associated boundary region of $\mathcal E_{R_Q}$. }
    \label{fig:pt}
\end{figure}
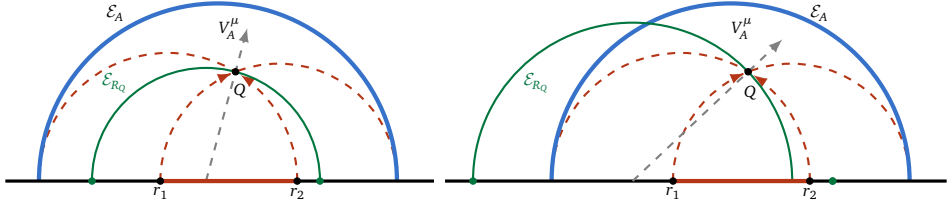

Let us  prove the statement for static intervals in the following:
\begin{itemize}
 \item 
Provided that, for any region $A$ the \emph{bit thread flow} $V_A^{\mu}$ is normal to the RT surface and satisfy $|V_A|=1/4G$ on the RT surface, then we have $|V_A|<1/4G$ away from RT surface.
\end{itemize}
The RT surfaces are static semicircles in pure AdS$_3$ spacetime. As is shown in Fig.\ref{fig:pt}, for a bulk point $Q$ inside the entanglement wedge $\mathcal{W}_{A}$, only the PEE threads emanating from the red region $[r_1,r_2]$ flows outside the $\mathcal{W}_{A}$ and contribute non-trivially to $V_A^{\mu}(Q)$. The vector $V_A^{\mu}(Q)$ determines a RT surface $\mathcal{E}_{R_Q}$ of a region $R_{Q}$, which passes through $Q$ and normal to $V_A^{\mu}(Q)$. 
Generally, there are following two possibilities:
\begin{enumerate}
\item  The whole red region lies inside $R_Q$, i.e. $[r_1,r_2]\subset R_Q$.
    \item 
    Only a part of the red region lies inside $R_Q$.
\end{enumerate}
The PEE threads from the red region $[r_1,r_2]$ and inside $R_Q$ 
will contribute positively to $V_A^{\mu}(Q)$ while those from the red region  but outside $R_Q$ contribute negatively to $V_A^{\mu}(Q)$. 

On the other hand, let us consider the region $R_{Q}$ where the \emph{bit thread flow} $V_{R_{Q}}(Q)$ has the same direction as $V_{A}(Q)$ and $|V_{R_{Q}}(Q)|=\frac{1}{4G}$. All the \emph{PEE thread flow} $V_{\textbf x}$ for $R_{Q}$ contribute positively along the direction of $V_{A}(Q)$. This means the contribution from the red region $[r_1,r_2]$  that lies inside $R_Q$ to $V_{A}(Q)$ is less than $\frac{1}{4G}$. 
As we have shown that the red region that lies outside $R_Q$ contribute negatively to $V_A(Q)$, we can conclude that $|V_{A}(Q)|<\frac{1}{4G}$. A similar argument also applies to the case where the field point lies outside $\mathcal{E}_{A}$, hence the statement is proved.

\section{PEE threads for a strip region}\label{app:strip}
Now let us construct the \emph{bit thread flow} $V_A^\mu$ for $(d-1)$-dimensional strip region with 
\begin{equation}
    x_1\in[-R,R],\quad x_i\in (-\infty,\infty), \text{for}\ i= 2,...,d-1.
\end{equation}
And the corresponding RT surface is given by \cite{Hubeny:2012ry}
\begin{equation}\label{strip-rt}
    \pm r(z)=\frac{z_*\sqrt\pi}{2(d-1)}\frac{\Gamma(\frac12+\frac{1}{2(d-1)})}{\Gamma(1+\frac{1}{2(d-1)})}
    -\frac{z}{d}\left(\frac{z}{z_*}\right)^{d-1}\ _2F_1\left(
    \frac12,\frac12+\frac{1}{2(d-1)},\frac32+\frac{1}{2(d-1)},\frac{z^{2(d-1)}}{z_*^{2(d-1)}}
    \right),
\end{equation}
where $z_*$ is the deepest point of minimal surface in the bulk, which reads
\begin{equation}
    z_*=\frac{2R(d-1)}{\sqrt\pi}\frac{\Gamma(1+\frac{1}{2(d-1)})}{\Gamma(\frac12+\frac{1}{2(d-1)})}.
\end{equation}
Due to the translational symmetry along $x_i$-direction, these $x_i$-components of $V_A^\mu$ vanishes and $V_A^\mu$ also enjoy this translational symmetry along $x_i$-direction. 
Thus it is enough to evaluate the field point at $(\bar{x}_1,\bar z)$ with $\bar{x}_{i\geq2}=0$.
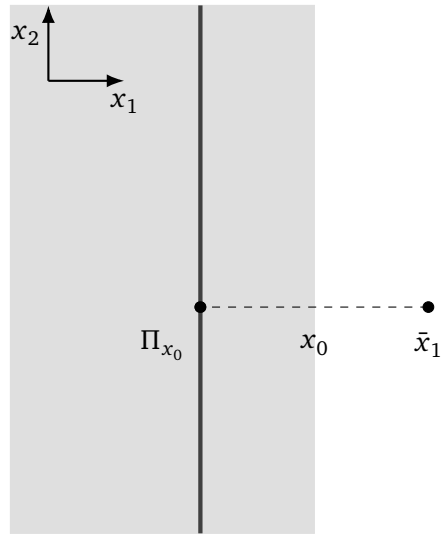
\begin{figure}
    \centering
    \begin{tikzpicture}
    \filldraw[lightgray!50] (-2-3-0.5,-3) rectangle (2-3-0.5,4);
    \filldraw[black] (0,0) circle (2pt);
    \filldraw[black] (-3,0) circle (2pt);
    \draw[darkgray,ultra thick](-3,-3)--(-3,4) ;
    \draw[dashed](0,0)--(-3,0);
    \filldraw[black] (-3,0) circle (2pt);
    \draw[] (-1.5,-0.5) node{$x_0$};
\draw[] (-3.5,-0.5) node{$\Pi_{x_0}$};
    \draw [black,thick,-Latex] (-5,3)--(-5,4);
    \draw [black,thick,-Latex] (-5,3)--(-4,3);
    \draw[] (-4,2.7) node{$x_1$};
    \draw[] (-5.3,3.6) node{$x_2$};
    \draw[] (0,-0.5) node{$\bar x_1$};
\end{tikzpicture}
    \caption{The blue-shaded region is a strip with the width $2R$ centered at the origin. 
    The strip is foliated by layers of thin plates $\Pi_{x_0}$ along the $x_1$-direction and $x_0$ is distance between the plate and the field point.
    }
    \label{fig:high-d,strip,plate}
\end{figure}
To get the \emph{bit thread flow} $V^\mu_A$, let us follow the similar trick as the sphere case.
We first foliate the strip into layers of thin plates $\Pi_{x_0}$ along $x_1$-direction (see Fig.\ \ref{fig:high-d,strip,plate} for an illustration), where $x_0$ is the distance between the field point and the plate.
Then we use \eqref{vx=0-HD} to determine the contribution from each thin plate to $V_A^\mu$ and integrate over all the plates to get the total $V_A^\mu$.
The results are 
\begin{equation}\label{eq:vax1,strip}
    V^{x_1}_A(\bar x_1,\bar z)
    =\frac{\bar z^{d+1}}{4G_N}\left(\frac{1}{\left((R-{\bar x_1})^2+\bar z^2\right)^{d/2}}-\frac{1}{\left((R+{\bar x_1})^2+\bar z^2\right)^{d/2}}\right),
\end{equation}
\begin{equation}\label{eq:vaz,strip}
    V_A^z(\bar x_1,\bar z)=\frac{\bar z^{d}}{4G_N}\left(
    \frac{\bar x_1+R}{((R+\bar x_1)^2+\bar z^2)^{d/2}}-\frac{\bar x_1-R}{((\bar x_1-R)^2+\bar z^2)^{d/2}}
    \right).
\end{equation}
One could check that this extrapolates to $d=2$ to recover \eqref{eq:2d,vaz}.

Unlike the spherical regions, however, the \emph{bit thread flow} for strip  $|V_A|<1/4G_N$ on the RT surface \eqref{strip-rt} and thus is not a bit thread configuration.


\end{document}